\documentclass[12pt]{article}
\usepackage{amssymb}
\usepackage{amsmath}
\usepackage{graphicx}

\usepackage{morefloats}

\begin{document} %


\title{Muon pair production via photon-induced scattering at the CLIC
in models with extra dimensions}


\author{
S. C. \.{I}nan\thanks{Electronic address: sceminan@cumhuriyet.edu.tr} \\
{\small Department of Physics, Sivas Cumhuriyet University,
58140, Sivas, Turkey} \\
A.V. Kisselev\thanks{Electronic address: alexandre.kisselev@ihep.ru} \\
{\small A.A. Logunov Institute for High Energy Physics, NRC
``Kurchatov Institute''}, \\
{\small 142281 Protvino, Russian Federation}
}

\date{}

\maketitle

\begin{abstract}
The photon-induced dimuon production $e^+e^- \rightarrow e^+ \gamma
\gamma e^- \rightarrow e^+ \mu^+\mu^- e^- $ at the CLIC is studied
in the framework of three models with extra dimensions. The electron
beam energies 750 GeV and 1500 GeV are considered. The total cross
sections  are calculated depending on the minimal transverse momenta
of the final muons. The sensitivity bounds on the parameters of the
models are obtained as functions of the CLIC integrated luminosity.
\end{abstract}

{\noindent Keywords: theories with extra dimensions, Kaluza-Klein
gravitons, CLIC collider; dimuon production in $e^+e^-$ collision,
photon-induced scattering. 
arXiv:1907.12824}


\section{Introduction} %
\label{sec:intr}

The  Standard Model (SM) has been validated by existing experiments
including the LHC date and has passed many tests very successfully
at the electroweak energy scale. However, many issues remain open in
the SM. One of the most fundamental of these problems is the
hierarchy problem. This open question involves the large energy gap
between the electroweak scale and the gravity scale. One of the
fundamental approaches to the solution of the hierarchy problem is
the theories that suggest the existence of extra dimensions (EDs).
Recently, many articles have been published on these theories, which
attracted great attention.

Scientists expect the LHC to elucidate much unanswered physics
problems. Nevertheless, this type of collider enables precision
measurements due to the nature of the proton-proton collisions.
Whereas, interactions of the electrons and positrons with
high-luminosity can provide higher precision than the proton-proton
interaction with too much background. Compact Linear Collider (CLIC)
is one of the most qualified $e^{+}e^{-}$ colliders. CLIC includes
normal conducting accelerating cavities and two-beam acceleration
\cite{bra}. It is used in a novel two-beam acceleration technique.
In this way, accelerating gradients could be obtained as $100$ MV/m.
The emission of beamstrahlung and production of background particles
at the CLIC is limited to an acceptance level by using flat beams
with the small vertical size \cite{CLIC_beam_str}. It is expected
that the CLIC detector will have the high transverse momentum
resolution for high-momentum muon tracks (see, for instance,
Fig.~7(a) in \cite{CLIC_beam_str}). One of the main requirements for
the CLIC detector is the overall detector coverage (hermeticity),
particularly in the forward region \cite{CLIC_herm}. This is needed
for lepton identification and missing energy. For particle energy
$7.5$ GeV the muon identification efficiency for the CLIC detector
was estimated to be about 99\% \cite{dan}.

To work CLIC at maximum efficiency, three energy stages are planned
\cite{bur}. The first one is at $\sqrt{s}=380$ GeV and can reach the
integrated luminosity $L=1000$ fb$^{-1}$. This era covers Higgs
boson, top and gauge sectors. It is possible to search for such SM
particles with the high precision \cite{dan}. Second operation is at
$\sqrt{s}=1500$ GeV. This stage is the highest center-of-mass energy
available with a single CLIC drive beam complex. In the second
stage, CLIC can give clues to beyond the SM physics. Moreover,
particular Higgs properties such as the Higgs self-coupling and the
top-Yukawa coupling and rare Higgs decay channels could be studied
\cite{lin}. In this stage, the maximum integrated luminosity value is
$2500$ fb$^{-1}$. The last stage is that CLIC has reached its
maximum center-of-mass energy value $\sqrt{s}=3000$ GeV and
integrated luminosity value $L=5000$ fb$^{-1}$. It is possible to do
the most precise examinations of the SM. Moreover, it is enabled to
discovery beyond the SM heavy particles of mass greater than $1500$
GeV \cite{dan}. The CLIC potential for new physics is presented in
\cite{CLIC_BSM}.

At the CLIC, as with all linear accelerators, $e \gamma$ and $\gamma
\gamma$   interactions are possible. Such interactions can be formed
in two ways: Compton backscattering \cite{Compton_photon,cb1,cb2}
and photon-induced reactions \cite{WWA,wwa1,wwa2}. In photon-induced
reactions, $\gamma \gamma$ and $e \gamma$ interactions can occur
spontaneously, unlike the Compton backscattering process. Therefore,
photon-induced reactions are much more useful than the Compton
backscattering procedure search for new physics beyond the SM. This
type of interaction can be studied by the Weizs\"{a}cker-Williams
approximation (WWA). There are great advantages of using the WWA.
Numerical calculations can be easily performed using simple
formulas. In addition, this method is useful in experimental
searches. Because it allows us to determine events number for the
process $\gamma \gamma\rightarrow X $ approximately with use of the
$e^{-}e^{+}\rightarrow e^{-}Xe^{+}$ process \cite{WWS}. Moreover,
photon induced reactions have very clean backgrounds since these
reactions do not involve interference with weak and strong
interactions. There are many phenomenological and experimental
studies in the literature on photon-induced process \cite{exp1,
phe1, phe2, bil1, atag, bil2, kok1, kok2}.

In WWA, the photons have very small virtuality. Therefore, scattered
angels of the emitting photons from the path of the electron along the
actual beam trajectory should be tiny. In this approximation,
the photon spectrum in incoming electron with the energy $E$
is given by the formula \cite{WWA} (see also \cite{Ozguven:2017})%
\begin{eqnarray}\label{WW_photons}
\frac{dN}{dE_\gamma} = \frac{\alpha_{\mathrm{em}}}{\pi E} \left[ \frac{(1 - x
+ x^2/2)}{x} \ln \frac{Q_{\mathrm{max}}^2}{Q_{\mathrm{min}}^2} -
\frac{m_e^2 x}{Q_{\mathrm{min}}^2} \left( 1 -
\frac{Q_{\mathrm{min}}^2}{Q_{\mathrm{max}}^2} \right) \right] ,
\end{eqnarray}
where $x = E_\gamma/E$ is the energy fraction of the photon, $m_e$
is the electron mass, $\alpha_{\mathrm{em}}$ is the fine structure
constant, and
\begin{equation}\label{Q2_limits}
Q_{\mathrm{min}}^2 = \frac{m_e^2 x^2}{1 - x} \;, \quad
Q_{\mathrm{max}}^2 = 2 \mathrm{\ GeV}^2 \;.
\end{equation}
Note that the contribution of the photons with virtualities more
than $Q_{\mathrm{max}}^2$ is negligible \cite{wwa1}. On the other
hand,
\begin{equation}\label{Q2_max}
Q_{\mathrm{max}}^2 = Q_{\mathrm{min}}^2 + E^2(1 - x) \,\theta^2 \;,
\end{equation}
where $E$ is the beam energy, $\theta$ is the electron scattering
angle. The photon spectrum \eqref{WW_photons} is singular at $x
\rightarrow 0$. So, we have $\theta \ll 1$, and the validity of the
WWA is justified.

In the photon-induced collisions, the luminosity spectrum
$dL^{\gamma\gamma}/dW$ can be found with using WWA as follows,
\begin{eqnarray}
\label{efflum} \frac{dL^{\gamma\gamma}}{dW} =
\int\limits_{y_{\min}}^{y_{\max}} {dy \frac{W}{2y}
f_{1}(\frac{W^{2}}{4y}, Q^{2}_{1}) f_{2}(y,Q^{2}_{2})} \;,
\end{eqnarray}
where $f_i = dN/dE_{\gamma i}$, $i=1,2$,
\begin{equation}\label{y_limits}
y_{\min} = \frac{W^2}{4(E - m_e)} \;, \quad y_{\max} = E - m_e \;.
\end{equation}
The cross-section for the process $e^+e^- \rightarrow e^+ \gamma
\gamma e^- \rightarrow e^+ \mu^+\mu^- e^- $ is obtained by
integrating subprocess cross section $d\hat {{\sigma}}_{\gamma\gamma
\to \mu^+\mu^-}(W)$ over the photon luminosity spectrum
\begin{eqnarray}
\label{completeprocess} d\sigma = \int\limits_{W_{\min}}^{W_{\max}}
dW \,\frac{dL^{\gamma\gamma}}{dW} \,d\hat {{\sigma}}_{\gamma\gamma
\to \mu^+\mu^-}(W) \;,
\end{eqnarray}
where
\begin{equation}\label{W_limits}
W_{\min} = 2p_{\bot\min} \;, \quad W_{\max} = 2(E - m_e) \;,
\end{equation}
and $p_{\bot\min}$ is the minimal transverse momentum of the final
muons.

Precision measurements at the CLIC can be regarded as complementary
searches for new physics beyond the SM carried out at the LHC. In
the present paper, we examine the potential of the photon-induced
dimuon production at the CLIC in the framework of three different
models with EDs. Both flat and warped metrics of the space-time are
considered. Recently, we have studied the muon pair production in
the photon-induced process at the LHC \cite{Inan_Kisselev:2018}. The
LHC discovery limits on 5-dimensional gravity scale for the process
$pp \rightarrow p \gamma \gamma p \rightarrow p \mu^+\mu^- p$ have
been calculated. One of the main goals of the present paper is to
study the CLIC search limit for the process $e^-e^+ \rightarrow e^-
\gamma \gamma e^+ \rightarrow e^- \mu^+\mu^- e^+$, and to compare it
with the LHC search limit. We will show that the CLIC sensitivity to
the effects coming from EDs are noticeably larger than the
corresponding LHC bounds.

\section{Scenario with extra dimensions and flat metric} %
\label{sec:ADD}

One of the promising possibilities to go beyond the SM is to consider a
theory in a space-time with extra spatial EDs. Such an approach is
motivated by the (super)string theory \cite{Polchinski:98}. One of
the main goals of the theories with EDs is to explain the hierarchy
relation between the electromagnetic and Planck scales. In the model
proposed by Arkani-Hamed, Dimopolous, Dvali and Antoniadis
\cite{Arkani-Hamed:1998}-\cite{Hamed2:1998}, called ADD, this
relation looks like
\begin{equation}\label{hierarchy_relation_ADD}
\bar{M}_{\mathrm{Pl}}^2 = V_d \bar{M}_D^{d+2} \;,
\end{equation}
where $d$ is the number of EDs, $V_d = (2\pi R_c)^d$ is the volume
of compact EDs with the radius $R_c$, $\bar{M}_{\mathrm{Pl}} =
M_{\mathrm{Pl}}/\sqrt{8\pi}$ is the reduced Planck mass, and
$\bar{M}_D$ is the reduced fundamental gravity scale in $D=4+d$
dimensions, $\bar{M}_D = M_D/(2\pi)^{d/(d+2)}$. It is assumed that
the fundamental gravity scale $M_D$ is in the TeV region. The huge
gap between the Planck and TeV scales can be justified by the large
value of $V_d$ (so-called ``large EDs'').

The masses of the Kaluza-Klein (KK) gravitons in the ADD model are
given by the formula \cite{Arkani-Hamed:1998}-\cite{Hamed2:1998}
\begin{equation}\label{ADD_masses}
m_n = \frac{n}{R_c}, \quad n = \sqrt{n_1^2 + n_2^2 + \cdots n_d^2}
\;,
\end{equation}
where $n_i = 0, 1, \ldots$ ($i=1,2, \ldots d$). We see that in the
scenario with large EDs, the mass splitting $\Delta m_{KK} = 1/R_c$
is very small. Thus, the mass spectrum of the gravitons can be
regarded as continuous.

\section{Scenario with one warped extra dimension} %
\label{sec:RS}

The Randall-Sundrum (RS) scenario with one ED and warped metric is
based on the following background metric \cite{Randall:1999}
\begin{equation}\label{RS_background_metric}
\quad ds^2 = e^{-2 \sigma (y)} \, \eta_{\mu \nu} \, dx^{\mu} \,
dx^{\nu} - dy^2 \;,
\end{equation}
where $\eta_{\mu\nu}$ is the Minkowski tensor with the signature
$(+,-,-,-)$, and $y$ is a compactified extra coordinate. The
periodicity condition $y=y + 2\pi r_c$ is imposed, and the points
$(x_\mu,y)$ and $(x_\mu,-y)$ are identified. Thus, we obtain a model
of gravity in a slice of the AdS$_5$ space-time compactified to the
orbifold $S^1\!/Z_2$ with the size $\pi r_c$. Since this orbifold
has two fixed points, $y=0$ and $y=\pi r_c$, two branes can be put
at these points. They are called Planck and TeV brane, respectively.
All SM fields are assumed to live on the TeV brane.

The classical action of the RS scenario looks like
\cite{Randall:1999}
\begin{align}\label{action}
S &= \int \!\! d^4x \!\! \int_{-\pi r_c}^{\pi r_c} \!\! dy \,
\sqrt{G} \, (2 \bar{M}_5^3 \mathcal{R}
- \Lambda) \nonumber \\
&+ \int \!\! d^4x \sqrt{|g^{(1)}|} \, (\mathcal{L}_1 - \Lambda_1) +
\int \!\! d^4x \sqrt{|g^{(2)}|} \, (\mathcal{L}_2 - \Lambda_2) \;.
\end{align}
Here $G_{MN}(x,y)$ is the 5-dimensional metric, $M,N = 0,1,2,3,4$.
The quantities
\begin{equation}
g^{(1)}_{\mu\nu}(x) = G_{\mu\nu}(x, y=0) \;, \quad
g^{(2)}_{\mu\nu}(x) = G_{\mu\nu}(x, y=\pi r_c)
\end{equation}
are induced metrics on the branes, $\mu = 0,1,2,3$. $\Lambda$ is a
five-dimensional cosmological constant, while $\Lambda_1$ and
$\Lambda_2$ are tensions on the branes. $\mathcal{L}_1$
$\mathcal{L}_2$ are brane Lagrangians, and $G = \det(G_{MN})$,
$g^{(i)} = \det(g^{(i)}_{\mu\nu})$. From the RS action
\eqref{action} one gets 5-dimensional Einstein-Hilbert's equations
\begin{align}
6 \sigma'^2 (y) &= - \frac{\Lambda}{4 \bar{M}_5^3} \;,
\label{sigma_deriv_eq} \\
3\sigma''(y) &= \frac{1}{4 \bar{M}_5^3} \, [\Lambda_1 \, \delta(y) +
\Lambda_2 \, \delta(\pi r_c - y)] \;. \label{sigma_2nd_deriv_eq}
\end{align}
In what follows, the reduced 5-dimensional gravity scales will be
used, $\bar{M}_5 = M_5 /(2\pi)^{1/3}$. Let us underline that
equations \eqref{sigma_deriv_eq}, \eqref{sigma_2nd_deriv_eq} contain
\emph{only derivatives} of the function $\sigma(y)$ and that
equation \eqref{sigma_2nd_deriv_eq} is symmetric with respect to the
branes.

As it was shown in details in \cite{Kisselev:2016} (see also
\cite{Kisselev:2013}), a general solution of equations
\eqref{sigma_deriv_eq}, \eqref{sigma_2nd_deriv_eq} is given by
\begin{equation}\label{sigma_solution}
\sigma (y) = \frac{\kappa}{2} ( |y| - |y - \pi r_c | ) +
\frac{\kappa \pi r_c}{2} - C \;,
\end{equation}
where the parameter $\kappa$ with a dimension of mass defines a
five-dimensional scalar curvature $\mathcal{R}^{(5)}$, and $C$ is
$y$\emph{-independent} quantity.%
\footnote{The term $\kappa \pi r_c/2$ is introduced in
\eqref{sigma_solution} for convenience only.}
In addition, the following fine tuning
\begin{align}
\Lambda &= -24 \bar{M}_5^3\kappa^2 \;, \label{Lambda_fine_tuning}
\\
\Lambda_1 &= - \Lambda_2 = 12 \bar{M}_5^3 \kappa
\label{Lambdas_fine_tuning}
\end{align}
must be realized \cite{Kisselev:2016}. From now on, it will be
assumed that $\kappa
> 0$, $\pi\kappa \,r_c \gg 1$. Then the hierarchy relation is of the
form
\begin{equation}\label{hierarchy_relation_RS}
\bar{M}_{\mathrm{Pl}}^2  = \frac{\bar{M}_5^3}{\kappa} \,e^{2C}
\!\left( 1 - e^{-2\pi \kappa r_c} \right)\Big|_{\pi\kappa \,r_c \gg
1} = \frac{\bar{M}_5^3}{\kappa} \, e^{2C} \;.
\end{equation}
The interactions of the gravitons $h_{\mu\nu}^{(n)}$ with the SM
fields on the physical (TeV) brane are given by the effective
Lagrangian
\begin{equation}\label{Lagrangian}
\mathcal{L}_{\mathrm{int}} = - \frac{1}{\bar{M}_{\mathrm{Pl}}} \,
h_{\mu\nu}^{(0)}(x) \, T_{\alpha\beta}(x) \, \eta^{\mu\alpha}
\eta^{\nu\beta} - \frac{1}{\Lambda_\pi} \sum_{n=1}^{\infty}
h_{\mu\nu}^{(n)}(x) \, T_{\alpha\beta}(x) \, \eta^{\mu\alpha}
\eta^{\nu\beta} \;,
\end{equation}
were $T^{\mu \nu}(x)$ is the energy-momentum tensor of the SM
fields, and the coupling constant of the massive modes is
\begin{equation}\label{Lambda_pi}
\Lambda_\pi \simeq \bar{M}_{\mathrm{Pl}} \, e^{-\kappa \pi r_c} =
\left( \frac{\bar{M}_5^3}{\kappa} \right)^{\!\!1/2} \, \!\!\! e^{C
-\kappa \pi r_c} \;.
\end{equation}
The graviton masses $m_n$ are defined from the boundary conditions
imposed on wave functions of the KK excitations. They result in the
equation (see, for instance, \cite{Kisselev:2013})
\begin{equation}\label{masses_eq}
J_1 (a_{1n}) Y_1(a_{2n}) - Y_1 (a_{1n}) J_1(a_{2n}) = 0 \;,
\end{equation}
where $J_1(x)$ and $Y_1(x)$ are the Bessel functions of the first
and second kind, respectively, and the following notations are
introduced
\begin{equation}\label{a_in}
a_{1n} = \frac{m_n}{\kappa} \,e^{-C} \;, \quad a_{2n}  =
\frac{m_n}{\kappa} \,e^{\kappa \pi r_c - C} \;.
\end{equation}

By taking different values of $C$ in eq.~\eqref{sigma_solution}, we
come to quite different physical models within the framework with
the warped metric. In particular, for $C = 0$, we come to the
original RS1 model \cite{Randall:1999} with the hierarchy relation
\begin{equation}\label{hierarchy_relation_RS1}
\bar{M}_{\mathrm{Pl}}^2  = \frac{\bar{M}_5^3}{\kappa} \!\left( 1 -
e^{-2\pi \kappa r_c} \right)\Big|_{\pi\kappa \,r_c \gg 1} =
\frac{\bar{M}_5^3}{\kappa} \;.
\end{equation}
In order \eqref{hierarchy_relation_RS1} to be satisfied, one has to
put $\bar{M}_5 \sim \kappa \sim \bar{M}_{\mathrm{Pl}}$
\cite{Randall:1999}. The graviton masses, as one can see from
\eqref{masses_eq}-\eqref{a_in}, are given by the formula
\begin{equation}\label{RS1_masses}
m_n = x_n \kappa \, e^{-\kappa \pi r_c} \;, \quad n=1,2, \ldots \;,
\end{equation}
where $x_n$ are zeros of $J_1(x)$. The coupling constant
\eqref{Lambda_pi} will be of the order of one TeV, if we put $\kappa
r_c \simeq 11.3$. It is in agreement with our assumption $\pi\kappa
\,r_c \gg 1$. Then the lightest graviton resonance has a mass of
order one--few TeV.

Taking $C = \pi\kappa \,r_c$, we come to the RS-like model with a
small curvature (RSSC model). For the first time, it was studied in
\cite{Giudice:2005}, see also
\cite{Kisselev:2016}-\cite{Kisselev:2013},
\cite{Kisselev:2005}-\cite{Kisselev:2006}. In such a model, the
hierarchy relations take the form
\begin{equation}\label{hierarchy_relation_RSSC}
\bar{M}_{\mathrm{Pl}}^2  = \frac{\bar{M}_5^3}{\kappa} \!\left(
e^{2\pi \kappa r_c} - 1 \right)\Big|_{\pi\kappa \,r_c \gg 1} =
\frac{\bar{M}_5^3}{\kappa} \,e^{2\kappa \pi r_c} \;.
\end{equation}
It is thanks to the exponential factor in
\eqref{hierarchy_relation_RSSC} that the mass hierarchy can be
satisfied even for moderate values of $\bar{M}_5$ and $\kappa$. For
instance, this relation holds, if one puts $\bar{M}_5 \sim 1$ TeV,
$\kappa \sim 1$ GeV, and $\kappa \,r_c =10.2$. On the contrary, the
RS1 hierarchy relation \eqref{hierarchy_relation_RS1} does not admit
the parameters $\kappa$, $\bar{M}_5$ to lie in these region. The
mass spectrum of the KK gravitons in the RSSC model, as it follows
from \eqref{masses_eq}-\eqref{a_in}, is defined as
\begin{equation}\label{RSSC_masses}
m_n = x_n \kappa \;, \quad n=1,2, \ldots \;.
\end{equation}
Note that in the limit $\kappa \rightarrow 0$, the hierarchy
relation for the flat metric with one ED
\eqref{hierarchy_relation_ADD} is reproduced from
\eqref{hierarchy_relation_RSSC}
\begin{equation}\label{flat_hierarchy_relation}
\bar{M}_{\mathrm{Pl}}^2  = \bar{M}_5^3 V_1 \;,
\end{equation}
where $V_1 = 2\pi r_c$ is the volume of ED. At the same time,
$\Lambda_\pi \rightarrow \bar{M}_{\mathrm{Pl}}$, and $m_n
\rightarrow n/r_c$ \cite{Kisselev:2013}.

The interaction Lagrangian of the radion field $\phi(x)$ on the
visible brane looks like
\begin{equation}\label{Lagrangian_radion}
\mathcal{L}_{\mathrm{rad}} = \frac{1}{\sqrt{3}\Lambda_{\pi}} \,
T^{\mu}_ {\mu}(x) \, \phi(x) \;.
\end{equation}
In the RSSC model, the mass scale $\Lambda_{\pi}$ is given by
\cite{Kisselev:2007}
\begin{equation}\label{lambda_enum}
\Lambda_{\pi} = 100 \left( \frac{\bar{M}_5}{\mathrm{TeV}}
\right)^{\!\!3/2} \left( \frac{\mathrm{100 \ MeV}}{\kappa}
\right)^{\!\!1/2} \! \mathrm{TeV} \;.
\end{equation}
As it follows from Eqs.~\eqref{Lagrangian_radion},
\eqref{lambda_enum}, there is no problem with the radion field
$\phi$, since for $\kappa \sim 0.1\div1$ GeV, $\bar{M}_5\sim 1$ TeV,
the quantity $1/(\sqrt{3}\Lambda_{\pi})$ is very small, and,
consequently, the coupling of the radion to the SM fields is
firmly suppressed. Let us underline that the magnitude of all
``graviton'' cross-sections (with either real or virtual production
of the KK gravitons) is defined by the fundamental gravity scale
$\bar{M}_5$, which is a few TeV, not by the coupling $\Lambda_{\pi}$.

To conclude, from the point of view of a 4-dimensional observer, the
models with $C=0$ and $C=\kappa\pi r_c$ are \emph{quite different
physical models}. The experimental signature of the RS1 model is a
production of heavy resonances, while the signature of the RSSC
model is a deviation of cross-sections from SM predictions.

\section{Photon-induced dimuon production } %
\label{sec:dimuon}

Let us consider the subprocess $\gamma\gamma \rightarrow \mu^+\mu^-$
of the photon-induced dimuon production in $e^+e^-$ collision. It's
matrix element squared is the sum of electromagnetic, KK graviton
and interference terms \cite{Atag:2009}
\begin{equation}\label{M2_tot}
|M|^{2} = |M_{\mathrm{em}} +  M_{\mathrm{\mathrm{KK}}}|^{2} \;,
\end{equation}
where
\begin{align}
|M_{\mathrm{em}}|^{2} &= -2\,e^4 \left[ \frac{\hat{s} +
\hat{t}}{\hat{t}} +
\frac{\hat{t}}{\hat{s} + \hat{t}} \right] , \label{M_em} \\
|M_{\mathrm{KK}}|^{2} &= \frac{1}{4} |S(\hat{s})|^{2}
\left[-\frac{\hat{t}}{8}
(\hat{s}^{3}+2\hat{t}^{3}+3\hat{t}\hat{s}^{2} +
4\hat{t}^{2}\hat{s}) \right] , \label{M_KK} \\
M_{\mathrm{em}} M_{\mathrm{KK}}^* + M_{\mathrm{em}}^*
M_{\mathrm{KK}} &= -\frac{1}{4} \,e^2 \,\mathrm{Re}S(\hat{s})
\,\left[ \,\hat{s}^{2}+2\hat{t}^{2} + 2\hat{s}\,\hat{t} \,\right]
\;. \label{M_int}
\end{align}
The quantity $S(s)$ contains summation over $s$-channel massive KK
excitations which can be calculated without specifying process,
$\hat{s}$, $\hat{t}$ are Mandelstam variables of the subprocess
$\gamma\gamma \rightarrow \mu^+\mu^-$, and
$e^2=4\pi\alpha_{\mathrm{em}}$.

In the ADD model, this sum is given by%
\footnote{We use the definition of the graviton field of
\cite{Han:1999}: $g_{AB} = \eta_{AB} + \sqrt{2} \bar{M}_D^{-1-d/2}
h_{AB}$.}
\begin{equation}\label{S_ADD}
\mathcal{S}_{\mathrm{ADD}}(\hat{s}) =
\frac{2}{\bar{M}_{\mathrm{Pl}}^2} \sum_{n_1, \ldots n_d =
1}^{\infty} \frac{1}{\hat{s} - m_n^2 + i \varepsilon} \;,
\end{equation}
where the masses $m_n$ are defined by Eq.~\eqref{ADD_masses}. Since
the sum is infinite for $d \geqslant 2$, an ultraviolet procedure is
needed \cite{Hewett:1999}-\cite{Giudice:1999}. In the
Han-Lykken-Zhang (HLZ) convention \cite{Han:1999}, the sum of
virtual KK exchanges is replaced by the integral in variable $m_n$
with the ultraviolet cutoff $M_S$, that results in
\begin{equation}\label{S_HLZ}
\mathcal{S}_{\mathrm{HLZ}}(s) = \frac{8\pi s^{d/2-1}}{M_D^{d+2}}
\left[ 2iI(x) + \pi \right] ,
\end{equation}
where $x = M_S/\sqrt{s}$, and
\begin{equation}\label{I}
I(x) = \left\{
         \begin{array}{ll}
          - \displaystyle \sum_{k=1}^{d/2-1} \frac{x^{2k}}{2k} - \ln(x^2 - 1) \;,
& \ d = \mathrm{even} \\
          - \displaystyle \sum_{k=1}^{(d-1)/2} \frac{x^{2k-1}}{2k-1} + \frac{1}{2}\ln\frac{x+1}{x-1} \;,
& \ d = \mathrm{odd} \;.
         \end{array}
       \right.
\end{equation}
In what follows, we put $M_S = M_D$.

In the Hewett convention, sum \eqref{S_ADD} is replaced by
\cite{Hewett:1999}
\begin{equation}\label{S_Hewwet}
\mathcal{S}_{\mathrm{H}} = \frac{\lambda}{M_H^4} \;,
\end{equation}
where $M_H$ is the unknown mass scale, presumably of order $M_D$. The
exact relationship between scales $M_H$ and $M_D$ is not calculable
without knowledge of the full theory. The parameter $\lambda = \pm
1$ is taken in analogy with the standard parametrization for contact
interactions.

Note that in the Giudice-Rattazzi-Wells convention
\cite{Giudice:1999}
\begin{equation}\label{S_GRW}
\mathcal{S}_{\mathrm{GRW}} = \frac{16\pi i}{(d - 2)\Lambda_T^4} \;,
\quad d > 2 \;,
\end{equation}
where $\Lambda_T$ is a cutoff scale. We will not use this
approximation for $\mathcal{S}_{\mathrm{ADD}}(s)$ in our numerical
analysis.

In the RS scenario, the contribution of $s$-channel gravitons is
given by the sum
\begin{equation}\label{S_RSSC_def}
\mathcal{S}_{\mathrm{RS}}(\hat{s}) =  \frac{2}{\Lambda_{\pi}^2}
\sum_{n=1}^{\infty} \frac{1}{\hat{s} - m_n^2 + i \, m_n \Gamma_n}
\;.
\end{equation}
Here $\Gamma_n$ denotes the total width of the graviton with the KK
number $n$ and mass $m_n$ \cite{Kisselev:2006}
\begin{equation}\label{KK_widths}
\Gamma_n = \frac{\rho \,m_n^3}{\Lambda_\pi^2} \;,
\end{equation}
where $\rho = 0.09$.

In the RS1 model, taking into account that the KK resonances are
very heavy, we put
\begin{equation}\label{S_RS1}
\mathcal{S}_{\mathrm{RS1}}(s) = \frac{2}{\Lambda_{\pi}^2}
\sum_{n=1}^{4} \frac{1}{\hat{s} - m_n^2 + i \, m_n \Gamma_n} \;.
\end{equation}
The contribution from other resonances to the sum \eqref{S_RSSC_def}
is negligible.

In the RSSC model, graviton sum \eqref{S_RSSC_def} can be calculated
analytically \cite{Kisselev:2006}
\begin{equation}\label{S_RSSC}
\mathcal{S}_{\mathrm{RSSC}}(s) = - \frac{1}{4\bar{M}_5^3 \sqrt{s}}
\; \frac{\sin (2A) + i \sinh (2\varepsilon)}{\cos^2 \!A + \sinh^2 \!
\varepsilon } \;,
\end{equation}
where
\begin{equation}\label{A_epsilon}
A = \frac{\sqrt{s}}{\kappa} \;, \qquad \varepsilon  = 0.045 \left(
\frac{\sqrt{s}}{\bar{M}_5} \right)^{\!\!3} .
\end{equation}
As was already mentioned above, in the RSSC model the  KK graviton
exchanges should lead to the deviations of the cross-sections from
the SM predictions.

\section{Numerical analysis and results} %
\label{sec:num}

The main goal of this section is to calculate the deviations of the
cross-sections from the SM predictions in several models with
EDs and to estimate the CLIC 95\% C.L. search limit for the
photon-induced process $e^+e^- \rightarrow e^+ \gamma\gamma e^-
\rightarrow e^+\mu^+\mu^-e^-$. We assume 100\% efficiency in the
reconstruction and identification of the final muons. The expected
collision energy $\sqrt{s}$ of the CLIC is 380 GeV (1st stage), 1500
GeV (2nd stage) or 3000 GeV (3rd stage), with the integrated
luminosities for unpolarized beams to be equal to 1000 fb$^{-1}$,
2500 fb$^{-1}$, and 5000 fb$^{-1}$, respectively, as mentioned
above. Our numerical results have shown that for the same values of
the parameters of the models, the deviations from the SM are much
smaller for $\sqrt{s} = 380$ GeV. That is why we will present our
result for $\sqrt{s} = 1500$ GeV and $\sqrt{s} = 3000$ GeV only.

\subsection{ADD model} %

\begin{figure}[htb]
\begin{center}
\includegraphics[scale=0.60]{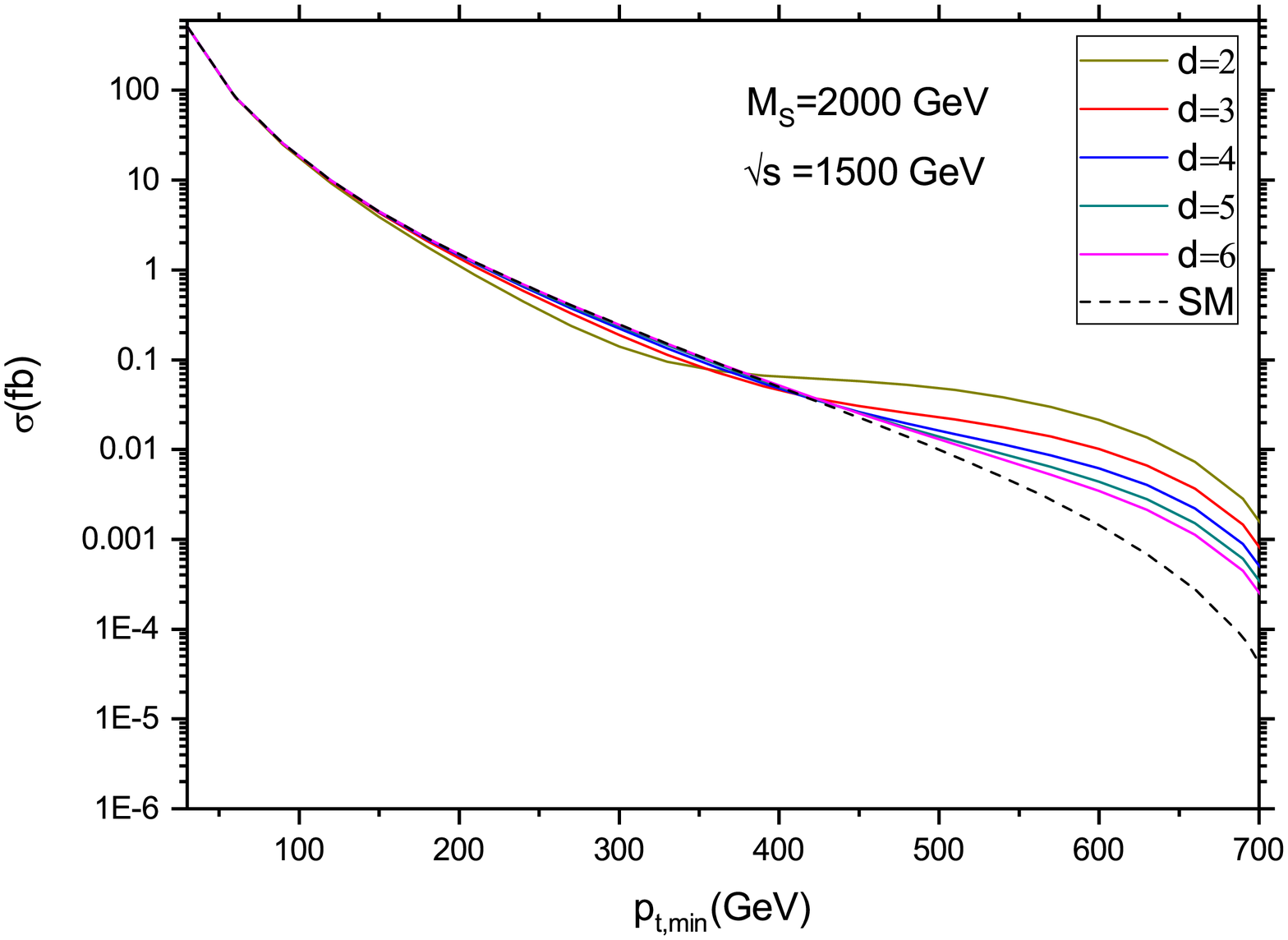}
\caption{The total cross-section for the process $e^+e^- \rightarrow
e^+ \gamma\gamma e^- \rightarrow e^+\mu^+\mu^-e^-$ in the ADD model
with the HLZ convention as a function of the muon transverse momenta
cutoff $p_{\mathrm{t,min}}$ for the CLIC invariant energy $\sqrt{s}
= 1500$ GeV and scale cutoff $M_S = 2$ TeV for different values of
the number of EDs. The dashed line denotes the SM contribution.}
\label{fig:HLZ_750_Ms_2.0}
\end{center}
\end{figure}
\begin{figure}[htb]
\begin{center}
\includegraphics[scale=0.60]{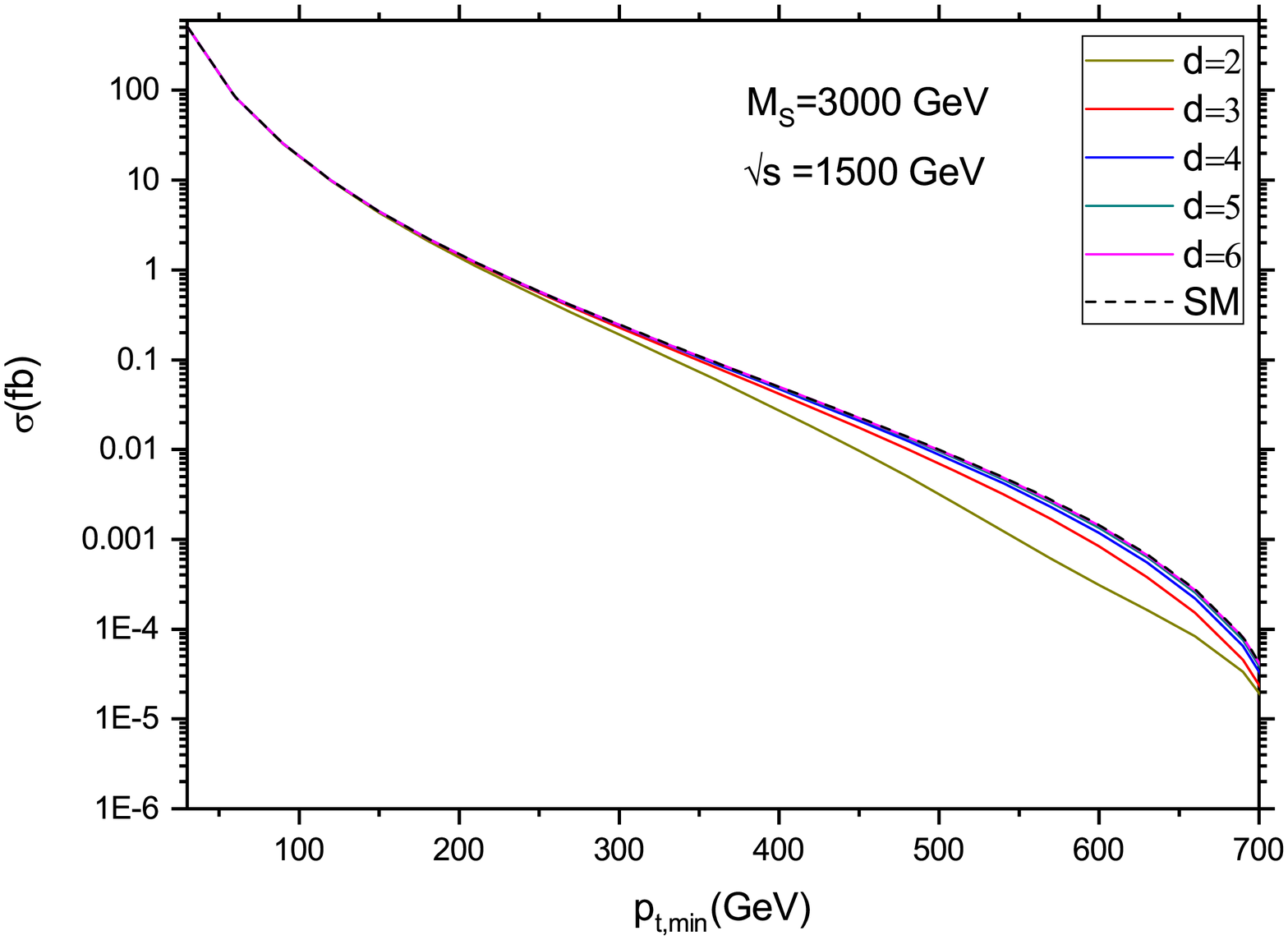}
\caption{The same as in Fig.~\ref{fig:HLZ_750_Ms_2.0}, but for $M_S
= 3$ TeV.} \label{fig:HLZ_750_Ms_3.0}
\end{center}
\end{figure}
\begin{figure}[htb]
\begin{center}
\includegraphics[scale=0.60]{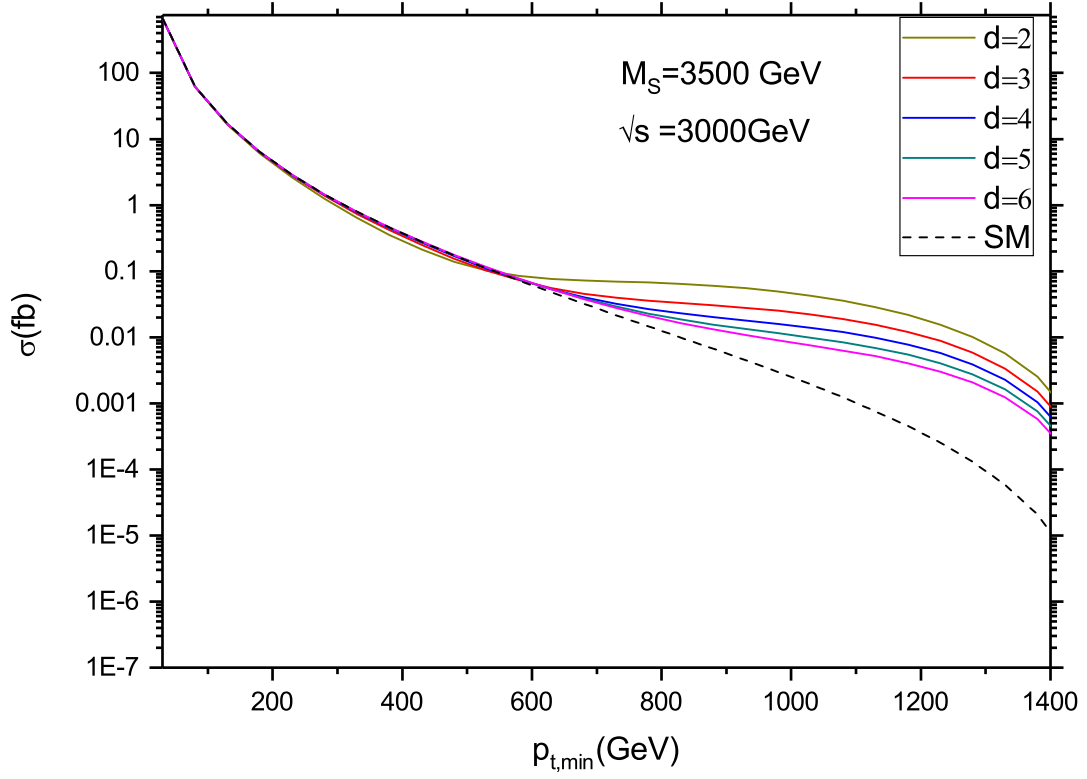}
\caption{The total cross section for the process $e^+e^- \rightarrow
e^+ \gamma\gamma e^- \rightarrow e^+\mu^+\mu^-e^-$ in the ADD model
with the HLZ convention as a function of $p_{\mathrm{t,min}}$ for
$\sqrt{s} = 3000$ GeV and scale cutoff $M_S = 3.5$ TeV for different
values of the number of EDs. The dashed line denotes the SM
contribution.} \label{fig:HLZ_1500_Ms_3.5}
\end{center}
\end{figure}
\begin{figure}[htb]
\begin{center}
\includegraphics[scale=0.60]{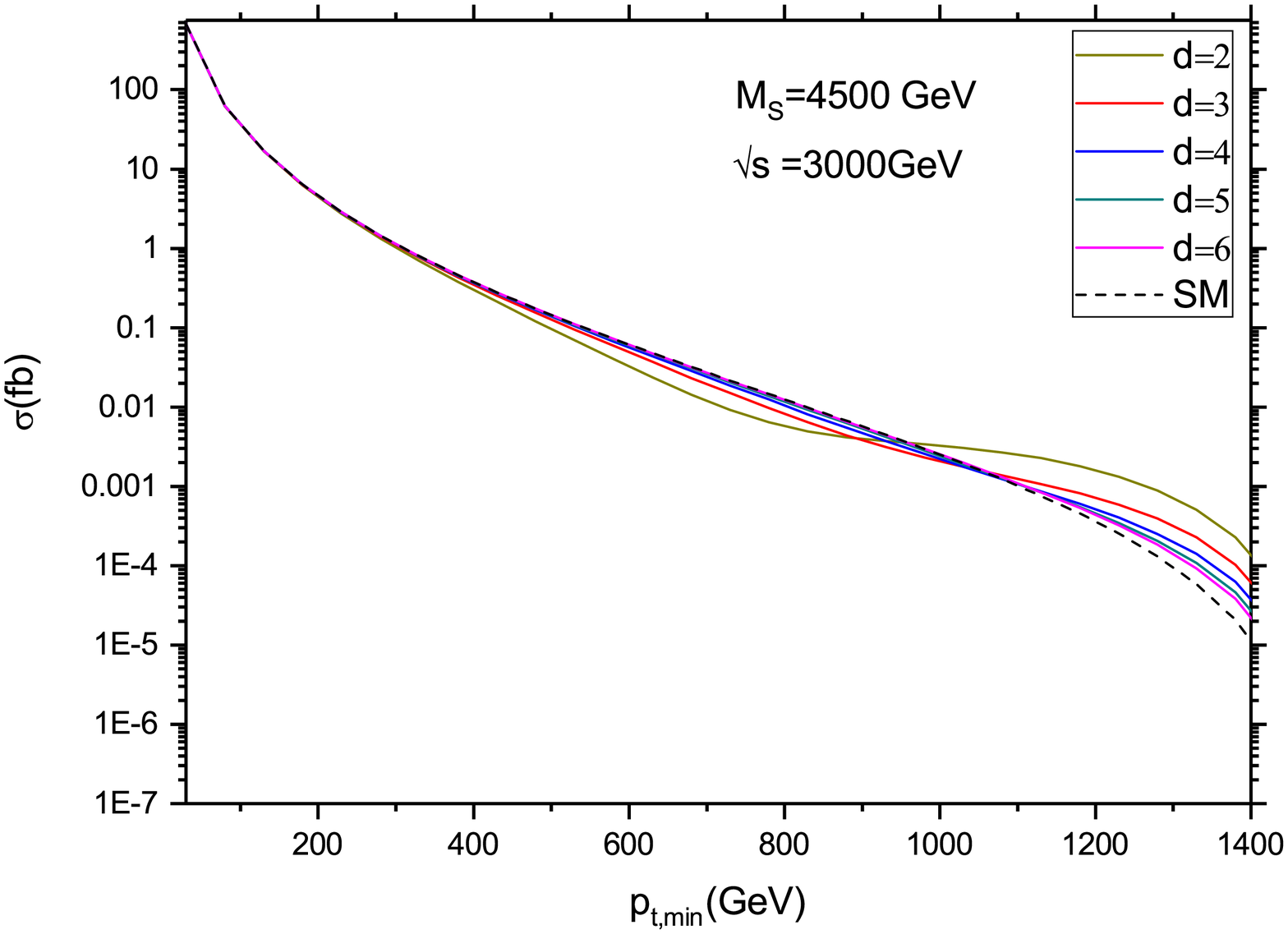}
\caption{The same as in Fig.~\ref{fig:HLZ_1500_Ms_3.5}, but for $M_S
= 4.5$ TeV.} \label{fig:HLZ_1500_Ms_4.5}
\end{center}
\end{figure}

In the ADD model with the Han-Lykken-Zhang (HLZ) convention
\cite{Han:1999} the invariant part of the subprocess $\gamma\gamma
\rightarrow \mu^+\mu^-$ is given by equations \eqref{S_HLZ},
\eqref{I}. The parameters are the number of EDs $d$ and the cutoff
scale $M_S$. The latter is believed to be of the order of the
$(4+d)$-dimensional gravity scale $M_D$. The results of our
calculations of the total cross-sections are shown in
Figs.~\ref{fig:HLZ_750_Ms_2.0}-\ref{fig:HLZ_1500_Ms_4.5} as
functions of the minimal transverse momenta of the final muons
$p_{\mathrm{t,min}}$. As one can see from these figures, the
deviations from the SM take place only for $p_{\mathrm{t,min}}
\gtrsim 200$ GeV for both energies. The curves correspond to
different values of  $d$ and $M_S$. As one can see, for $M_S = 2$
TeV and $M_S = 4.5$ TeV, the cross-sections have nontrivial
dependence on $d$. The interference term of the cross sections could
be relatively large in absolute value and negative for some mediate
$p_{t,\min}$ (150-450 GeV). As a result, in this $p_{t,\min}$ region
cross-section slightly rises with $d$ for some values of $M_S$, as
one can see in Figs.~\ref{fig:HLZ_750_Ms_2.0},
\ref{fig:HLZ_1500_Ms_4.5}.
%

%

We have calculated the statistical significance $SS$ using formula
\cite{SS}
\begin{equation}\label{SS_def}
SS = \sqrt{2[(S+B) \,\ln(1 + S/B) - S]} \;,
\end{equation}
where $S(B)$ is a number of the signal (background) events. Note
that $SS \simeq S/\sqrt{B}$ for $S \ll B$. We have assumed that the
uncertainty of the background is negligible.

The main contribution to the SM background comes from the SM process
$e^-e^+ \rightarrow e^- \gamma\gamma e^+ \rightarrow e^- \mu^+\mu^-
e^+$, going via subprocess $\gamma\gamma \rightarrow \mu^+\mu^-$.
Note that the other SM process $e^-e^+ \rightarrow e^- ZZ (WW) e^+
\rightarrow e^- \mu^+\mu^- e^+$ may not be taken into account as a
background, since as given in \cite{ZZ_lum}, the $Z$-$Z$ luminosity
function is 100 times smaller than the photon-photon luminosity
function.

It can be considered that the dimuon pair-production
$e^+e^-\rightarrow \mu^+\mu^-$ also contributes to the background
when the final-state electrons in the photon-induced processes are
almost along the beamlines, and they can not be detected. As it is
shown in Appendix~A, the total cross section of this process
$\sigma(p_t > 30 \mathrm{\ GeV})$ is very small with respect to our
main SM background. Moreover, the final state $\mu^+\mu^-$ in the
s-channel process $e^+e^- \rightarrow \mu^+\mu^-$ has a quite
different $p_t$ distribution. Hence, it can be easily discerned from
our process.

\begin{figure}[htb]
\begin{center}
\includegraphics[scale=0.75]{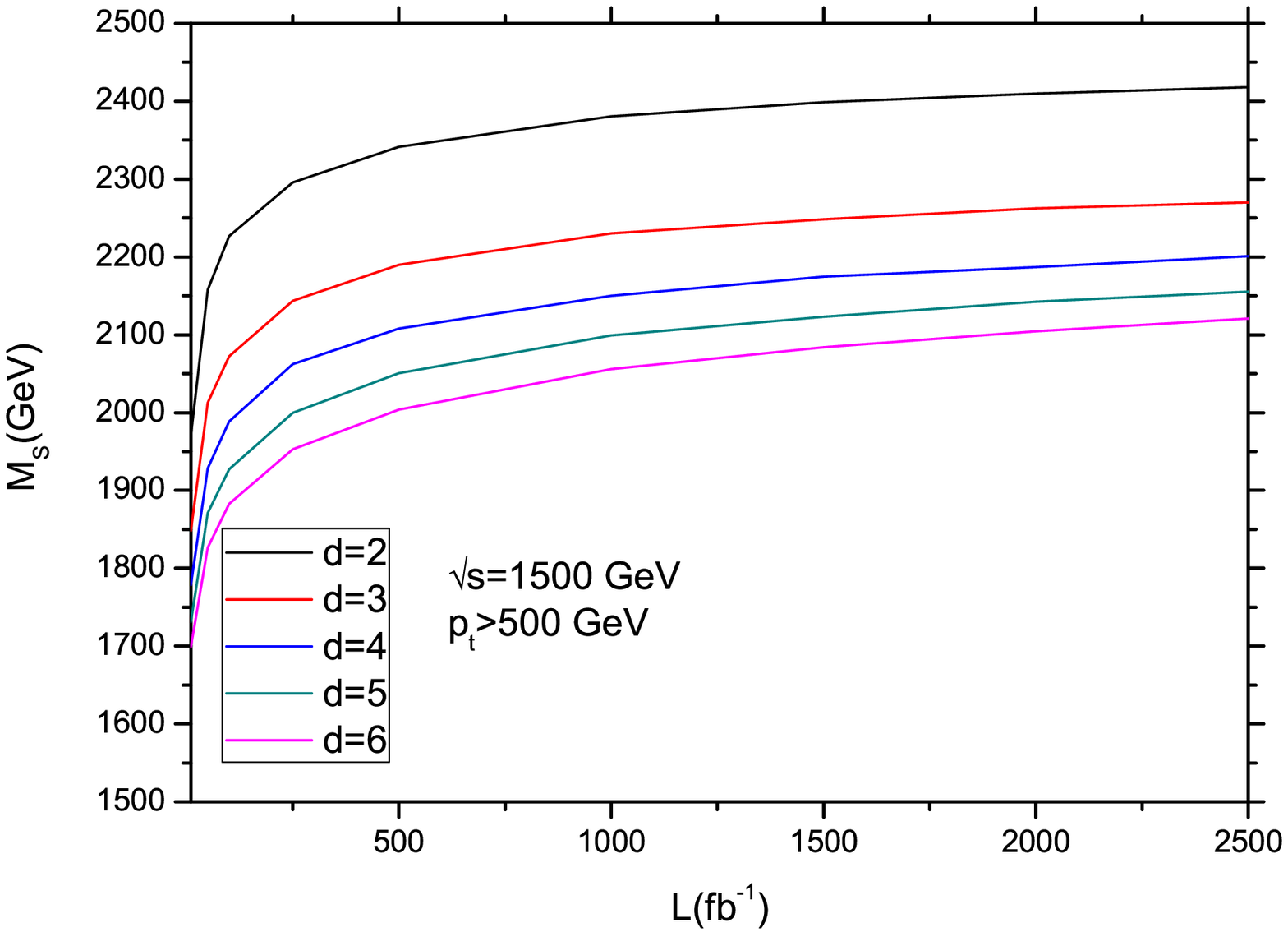}
\caption{The $95\%$ C.L. CLIC search bound in the ADD model with the
HLZ convention for $\sqrt{s} = 1500$ GeV, $p_t>500$ GeV as a
function of the integrated luminosity $L$.} \label{fig:S_HLZ_750}
\end{center}
\end{figure}

\begin{figure}[htb]
\begin{center}
\includegraphics[scale=0.75]{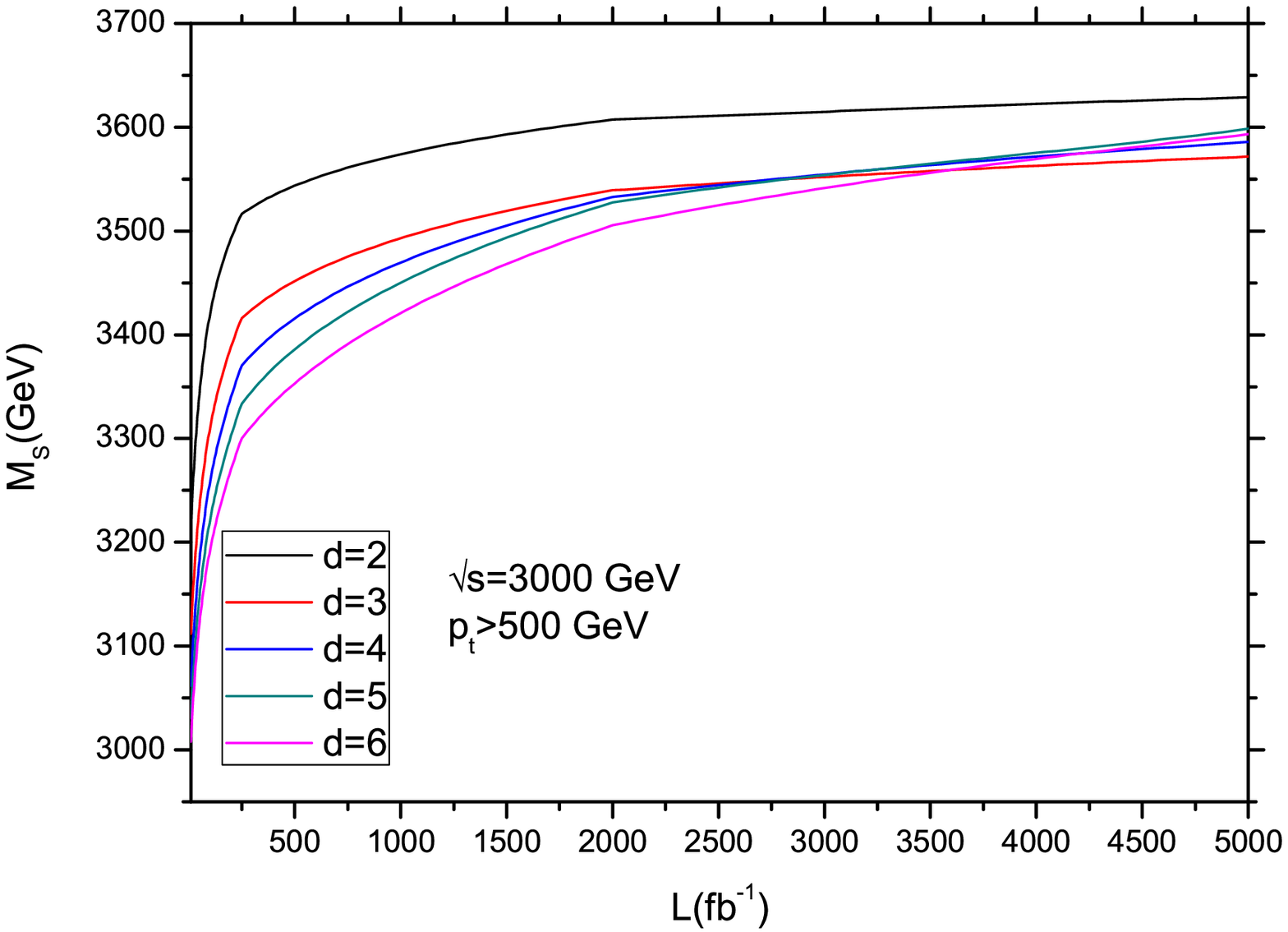}
\caption{The same as in Fig.~\ref{fig:S_HLZ_750}, but for $\sqrt{s}
= 3000$ GeV.} \label{fig:S_HLZ_1500}
\end{center}
\end{figure}

The 95\% C.L. search limits for two CLIC energies are presented in
Figs.~\ref{fig:S_HLZ_750}, \ref{fig:S_HLZ_1500} as functions of the
number of EDs and CLIC integrated luminosity for $\sqrt{s} = 1500$
GeV and $\sqrt{s} = 3000$ GeV. The bounds on $M_S$ are as large as
3629(3593) GeV for $d=2(6)$ can be achieved for the CLIC energy
$\sqrt{s} = 3000$ GeV and integrated luminosity  $L = 5000 \mathrm{\
fb}^{-1}$.
%

\begin{figure}[htb]
\begin{center}
\includegraphics[scale=0.65]{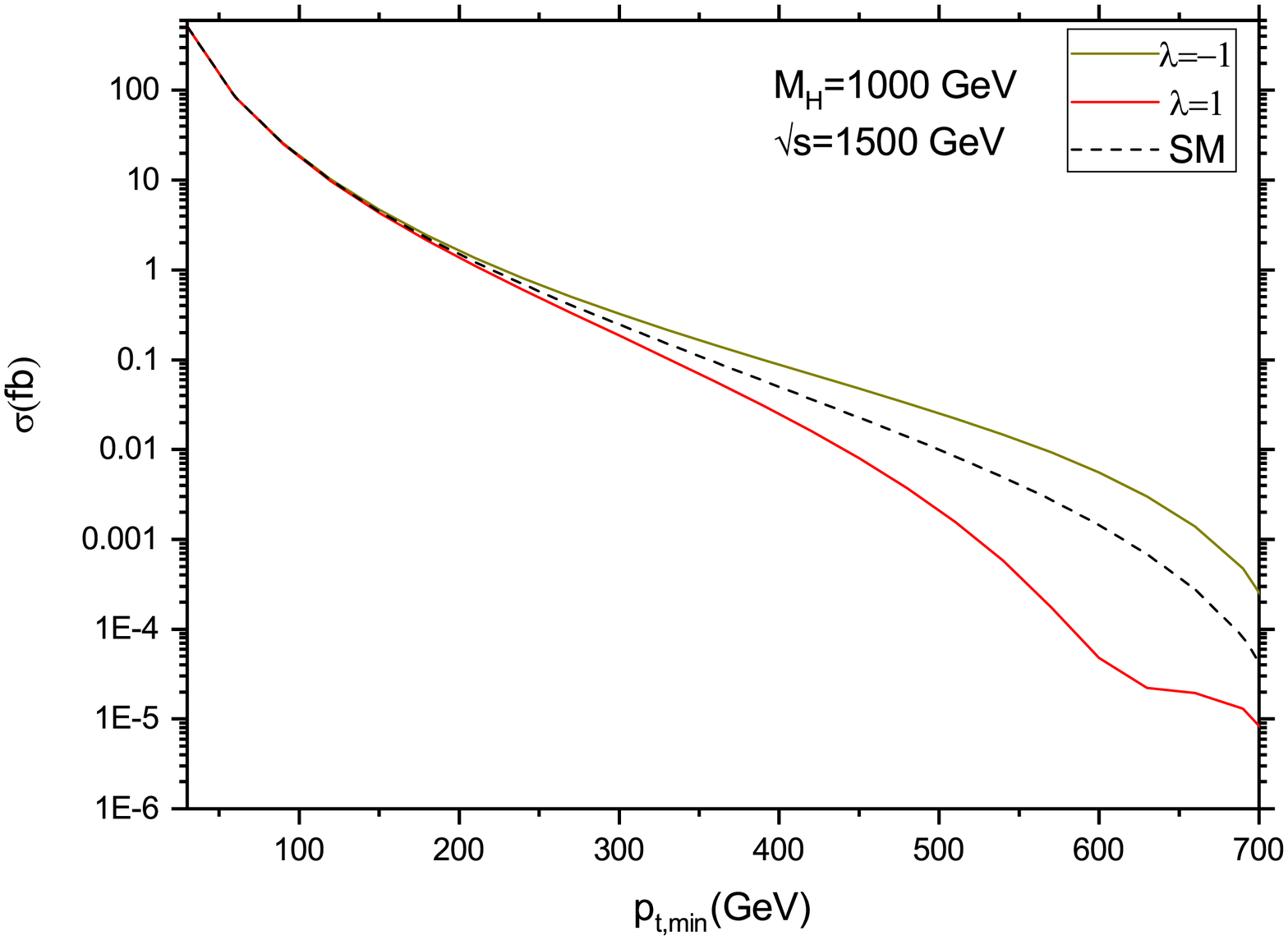}
\caption{The total cross-section for the process $e^+e^- \rightarrow
e^+ \gamma\gamma e^- \rightarrow e^+\mu^+\mu^-e^-$ in the ADD model
with the Hewett convention as a function of $p_{t,\mathrm{min}}$ for
$\sqrt{s} = 1500$ GeV.} \label{fig:H_750_Ms_2.0}
\end{center}
\end{figure}

\begin{figure}[htb]
\begin{center}
\includegraphics[scale=0.65]{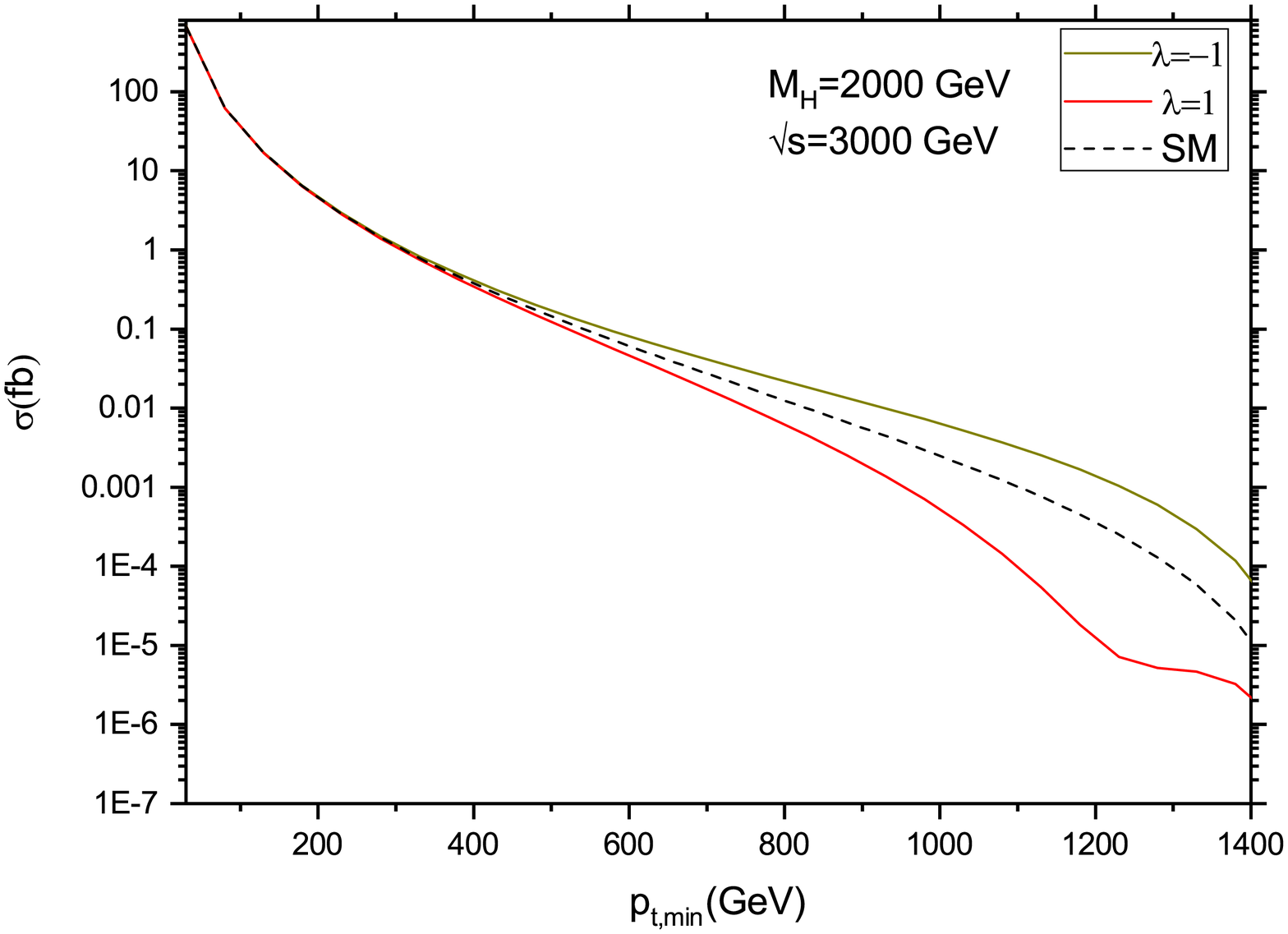}
\caption{The same as in Fig.~\ref{fig:H_750_Ms_2.0}, but for
$\sqrt{s} = 3000$ GeV and $M_S = 3.5$ TeV.}
\label{fig:H_1500_Ms_3.5}
\end{center}
\end{figure}

The cross-sections in the Hewett convention of the ADD model
\cite{Hewett:1999} depend on the ultraviolet cutoff $M_H$ and sign
of the parameter $\lambda$ in \eqref{S_Hewwet}. They are shown in
Figs.~\ref{fig:H_750_Ms_2.0}, \ref{fig:H_1500_Ms_3.5} as functions
of $p_{\mathrm{t,min}}$, both for positive and negative sign of the
parameter $\lambda$ in \eqref{S_Hewwet}. As one can see from these
figures, for $p_{\mathrm{t,min}} > 300$ GeV the SM and KK
contributions can be defined well.
%

\begin{figure}[htb]
\begin{center}
\includegraphics[scale=0.75]{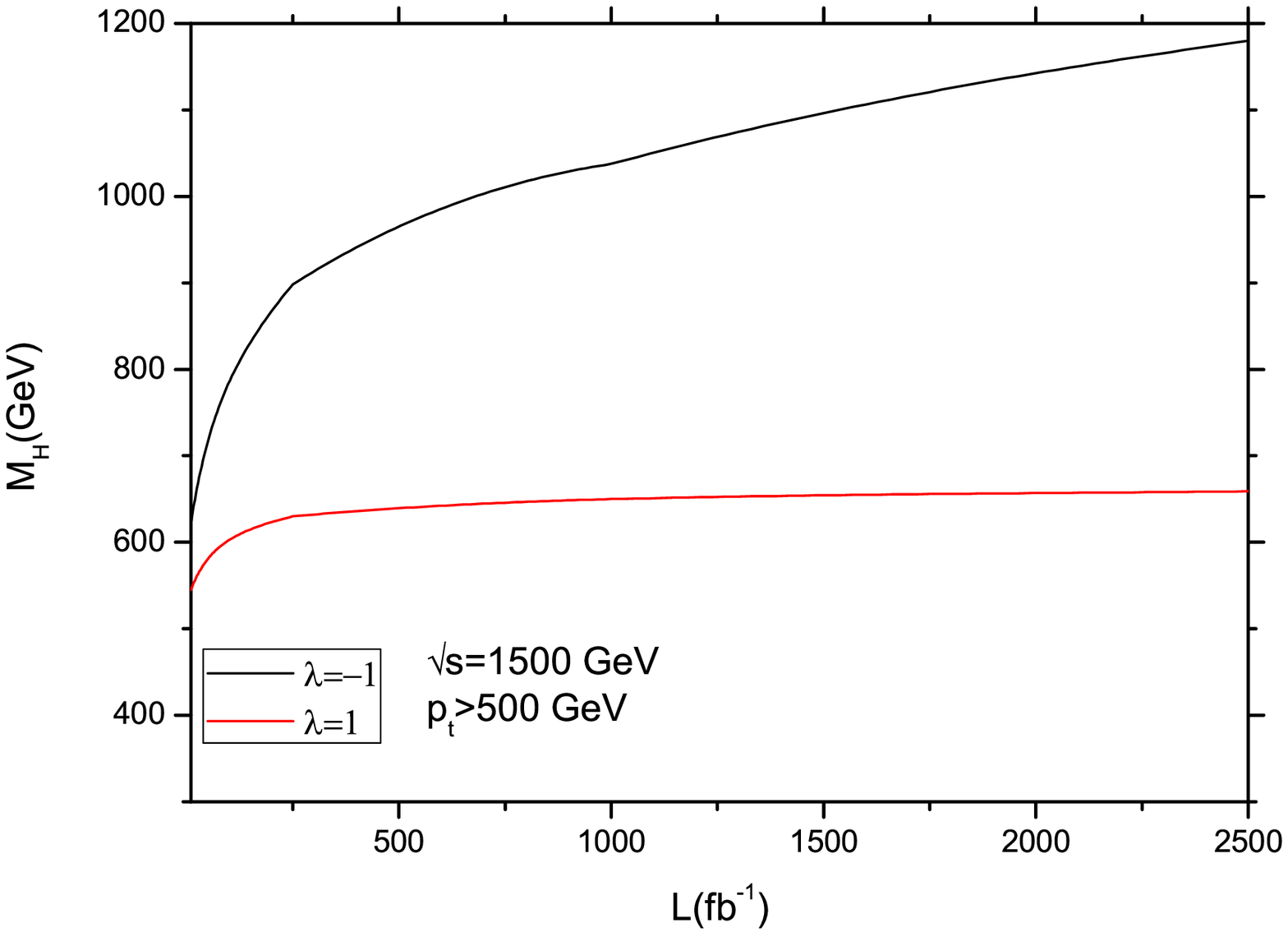}
\caption{The $95\%$ C.L. CLIC search bound in the ADD model with the
Hewett convention for $\sqrt{s} = 1500$ GeV, $p_t>500$ GeV as a
function of the integrated luminosity $L$.} \label{fig:S_H_750}
\end{center}
\end{figure}

\begin{figure}[htb]
\begin{center}
\includegraphics[scale=0.75]{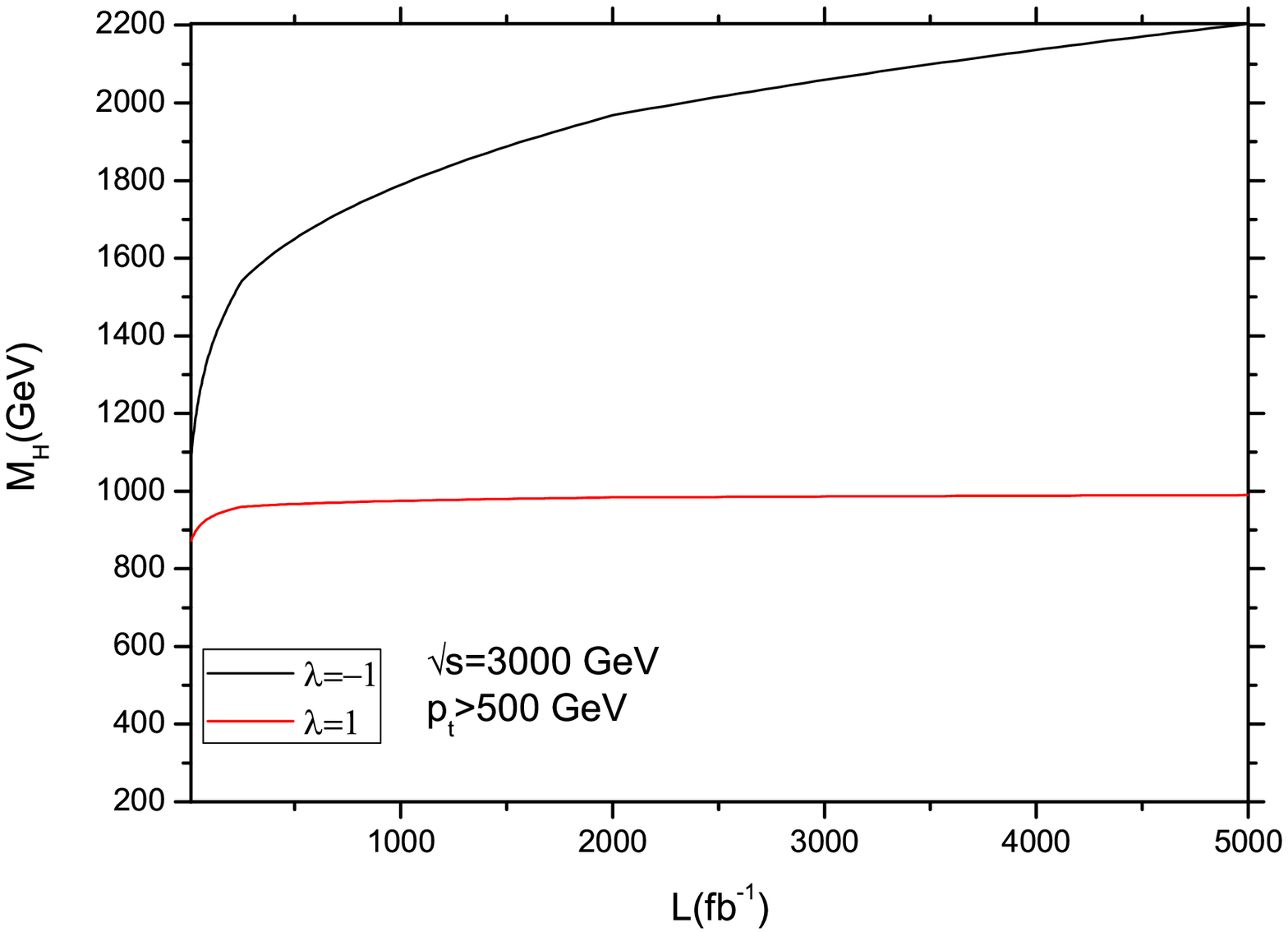}
\caption{The same as in Fig.~\ref{fig:S_H_750}, but for $\sqrt{s} =
3000$ GeV.} \label{fig:S_H_1500}
\end{center}
\end{figure}

The 95\% C.L. bounds on the cutoff scale $M_H$ as functions of the
CLIC integrated luminosity are given in Figs.~\ref{fig:S_H_750},
\ref{fig:S_H_1500}. They demonstrate us that the case $\lambda = -1$
is clearly preferable to the case $\lambda = 1$. The bound
$M_H=2204$ GeV can be achieved for $\sqrt{s} = 3000$ GeV, $L = 5000
\mathrm{\ fb}^{-1}$. Note that in the Hewett scheme there is no
dependence on the number of EDs. The bounds obtained should be
compared with the LHC bounds on the parameters in the ADD model
(see, for instance, \cite{CMS_bounds}).
%

\subsection{RS model} %

\begin{figure}[htb]
\begin{center}
\includegraphics[scale=0.60]{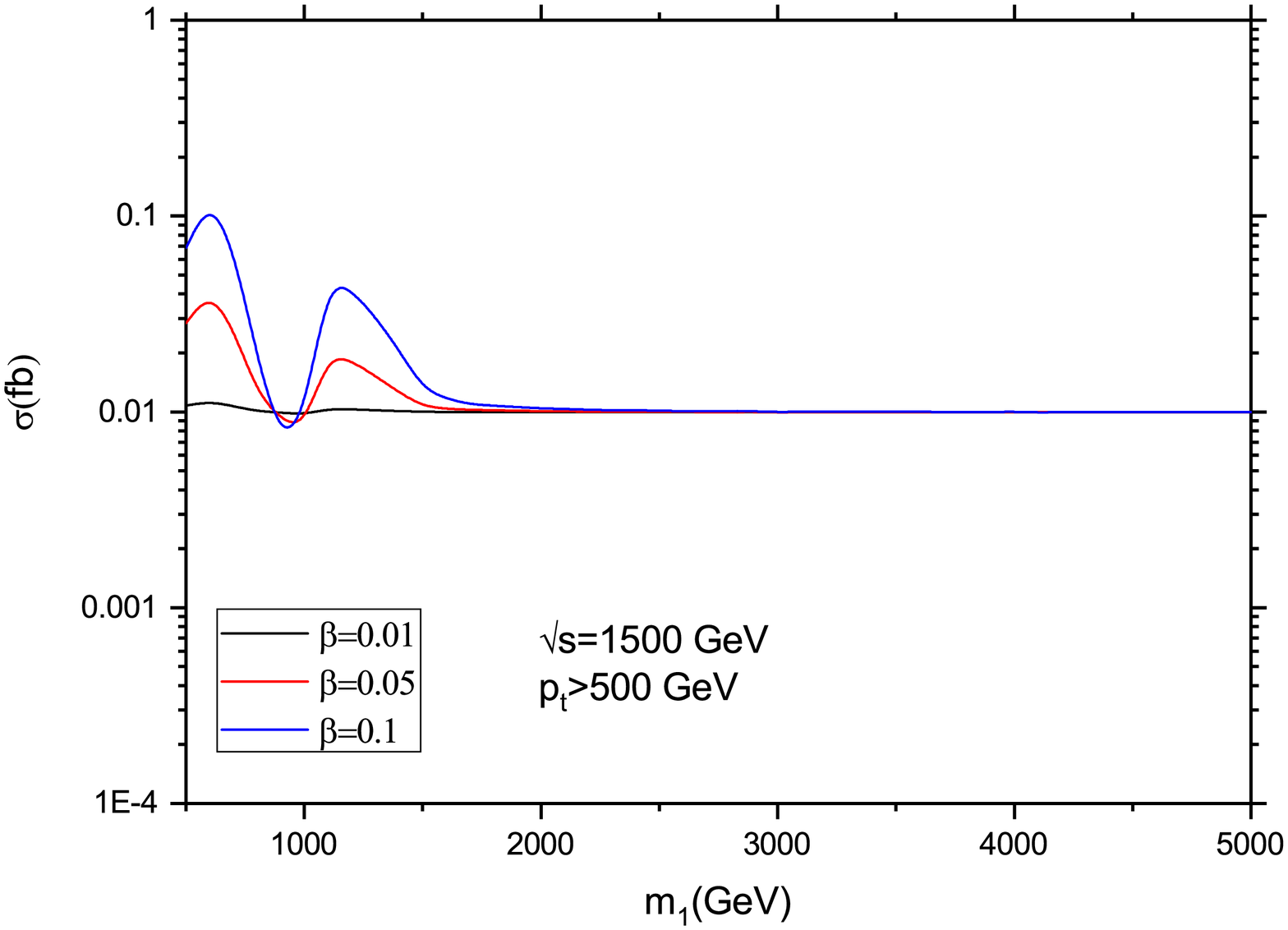}
\caption{The total cross-section for the process $e^+e^- \rightarrow
e^+ \gamma\gamma e^- \rightarrow e^+\mu^+\mu^-e^-$ in the RS model
as a function of the mass of the lightest KK resonance $m_1$ for
$\sqrt{s} = 1500$ GeV and different values of the parameter $\beta$.
The muon transverse momenta cutoff $p_t>500$ GeV. The dashed line
denotes the SM contribution.} \label{fig:RS_750_500}
\end{center}
\end{figure}

\begin{figure}[htb]
\begin{center}
\includegraphics[scale=0.60]{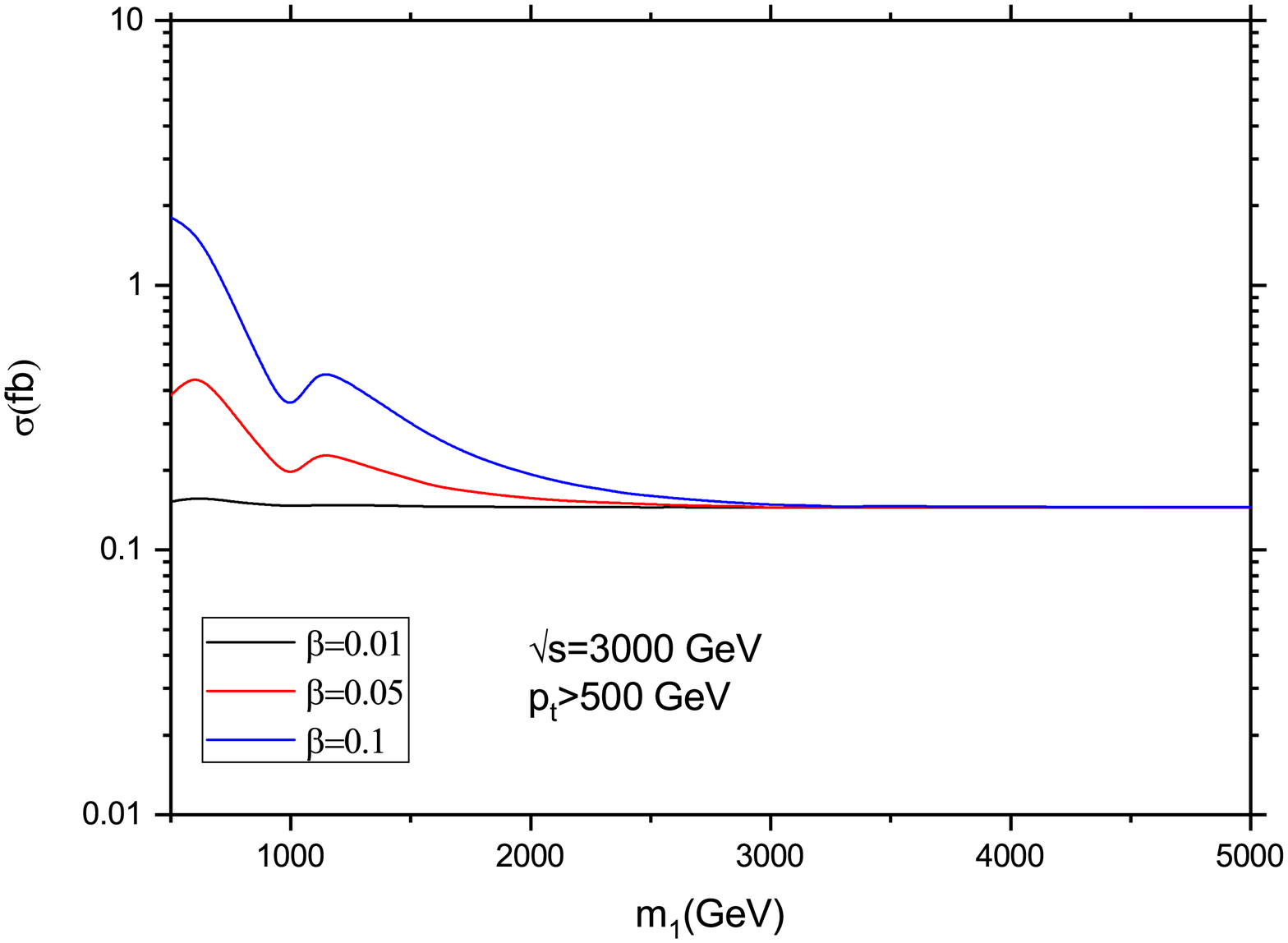}
\caption{The same as in Fig.~\ref{fig:RS_750_500}, but for $\sqrt{s}
= 3000$ GeV.} \label{fig:RS_1500_500}.
\end{center}
\end{figure}

We have also calculated the cross-sections in the Randall-Sundrum
model \cite{Randall:1999} using formula \eqref{S_RS1}. The
corresponding curves are shown in Figs.~\ref{fig:RS_750_500},
\ref{fig:RS_1500_500} as functions of the mass $m_1$ of the lightest
graviton for three values (0.01, 0.05, 0.1) of the ratio
\begin{equation}\label{beta}
\beta = \frac{\kappa}{\bar{M}_{\mathrm{Pl}}} \;.
\end{equation}
The oscillations of the curves in these figures correspond to the
resonance character of the invariant part of the $s$-channel
amplitude in the RS model \eqref{S_RS1}. Let us indicate on the
strong $\beta$-dependence of the cross section for $m_1 \lesssim 1.5
(2.0)$ TeV at $\sqrt{s} = 1500(3000)$ GeV. For $\beta = 0.01$ the
deviations from the SM are negligible for all $m_1$.
%

\begin{figure}[htb]
\begin{center}
\includegraphics[scale=0.60]{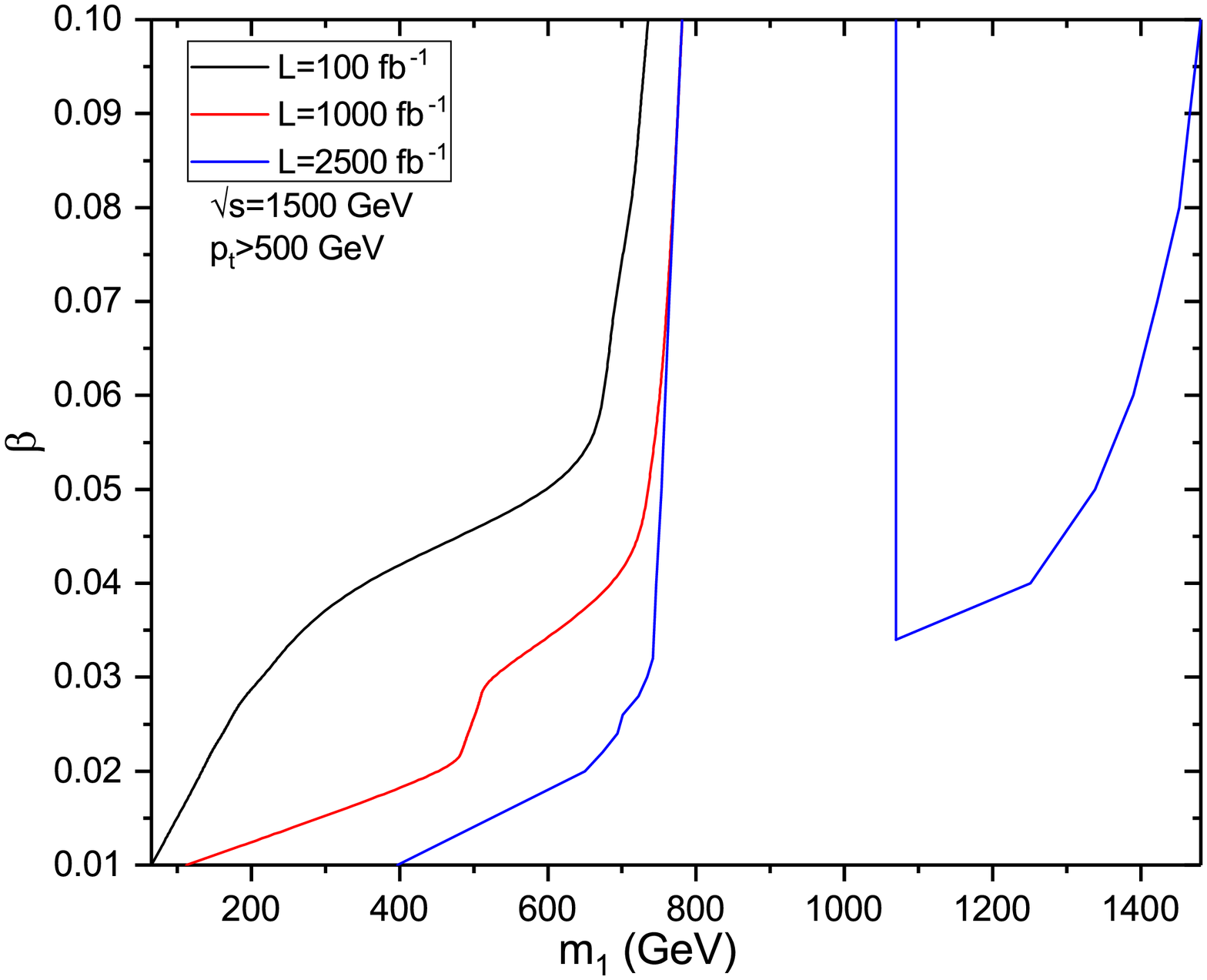}
\caption{The $95\%$ C.L. exclusion region for the parameters $m_1$
and  $\beta$ in the RS model for $\sqrt{s} = 1500$ GeV, $p_t>500$
GeV and three values of the CLIC integrated luminosity $L$.
The excluded regions are defined by the area over the curves.}
\label{fig:S_RS_750_500}
\end{center}
\end{figure}

\begin{figure}[htb]
\begin{center}
\includegraphics[scale=0.60]{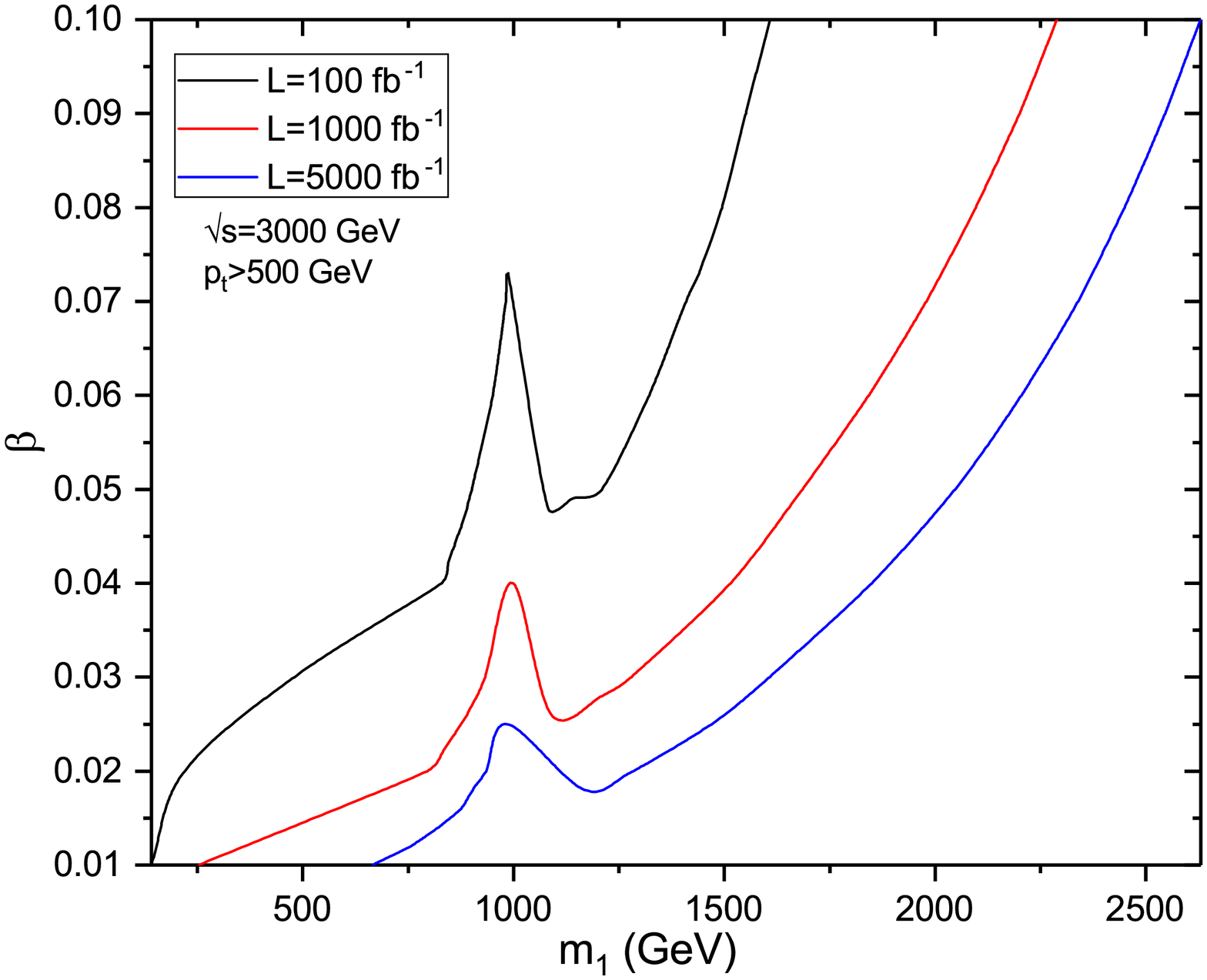}
\caption{The same as in Fig.~\ref{fig:S_RS_750_500}, but for
$\sqrt{s} = 3000$ GeV and different values of $L$.}
\label{fig:S_RS_1500_500}
\end{center}
\end{figure}

The estimation for the $95\%$ C.L. parameter exclusion region is
shown in Figs.~\ref{fig:S_RS_750_500},  \ref{fig:S_RS_1500_500} for
three expected values of the CLIC integrated luminosity. For both
energies, the value 500 GeV was taken as the minimal transverse
momenta of the final muons. Due to the oscillation behavior of the
total cross sections, their values are closed to the SM cross
sections for all $0.01 \leqslant \beta \leqslant 0.1$ in the mass
interval $m_1 = 750-1070$ GeV for $\sqrt{s}=1500$ GeV, see
Fig.~\ref{fig:RS_750_500}. As a result, the exclusion region has the
``hole'' in these mass region, as one can see in
Figs.~\ref{fig:S_RS_750_500}. As for invariant energy $\sqrt{s}=
3000$ GeV, similar oscillations of the cross section presented in
Fig.~\ref{fig:RS_1500_500} result in the ``bumps'' on the exclusion
curves in the $m_1 = 865-1170$ GeV region, see
Fig.~\ref{fig:S_RS_1500_500}.

The best lower bound which can be achieved, is equal to $m_1 = 2629$
GeV for $\beta=0.1$. The present experimental bounds on $m_1$ are
stronger \cite{LHC_BSM_bounds}. Thus, we don't expect that, instead
of the very high integrated luminosity of the CLIC 3rd stage, the
existing experimental bounds on $m_1$ could be improved in the
photon-induced dimuon production.
%

\subsection{RSSC model} %

\begin{figure}[htb]
\begin{center}
\includegraphics[scale=0.70]{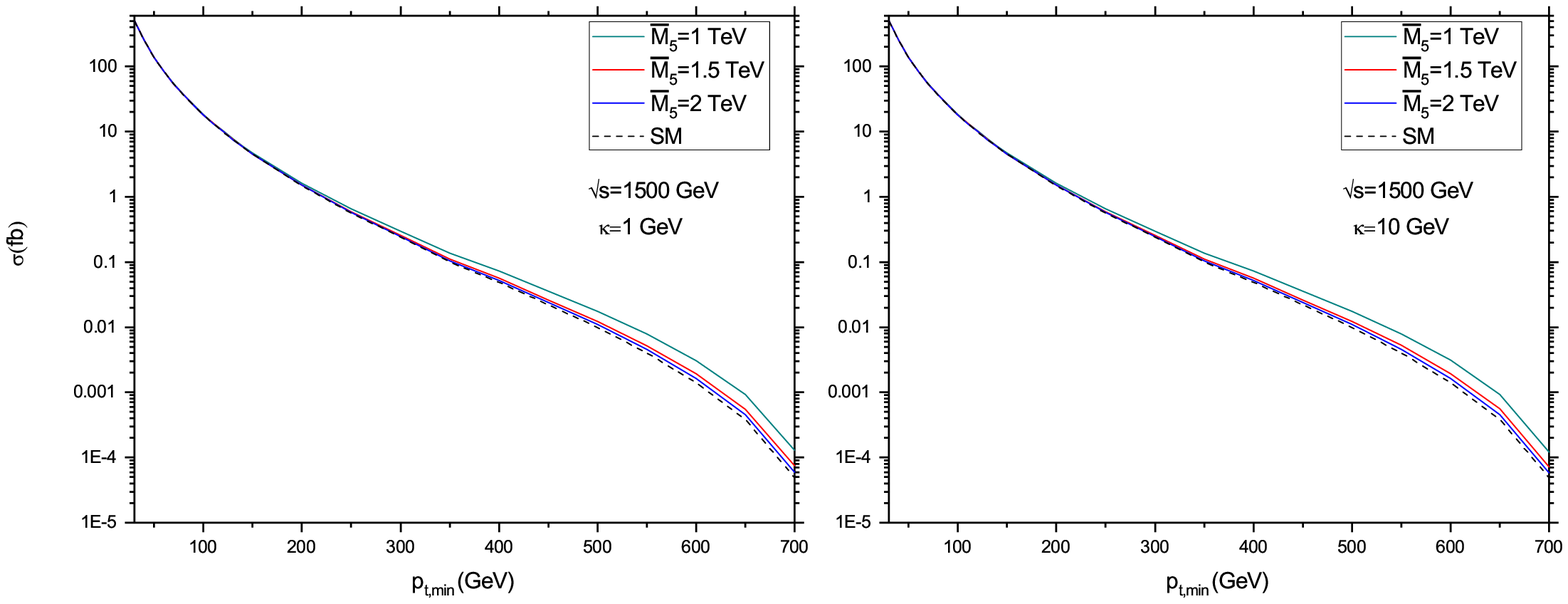}
\caption{The total cross-sections for the process $e^+e^-
\rightarrow e^+\mu^+\mu^-e^-$ in the RSSC model as a function of
$p_{t,\mathrm{min}}$ for $\sqrt{s} = 1500$ GeV and different values
of $\bar{M}_5$ and $\kappa$.} \label{fig:RSSC_750}
\end{center}
\end{figure}

\begin{figure}[htb]
\begin{center}
\includegraphics[scale=0.70]{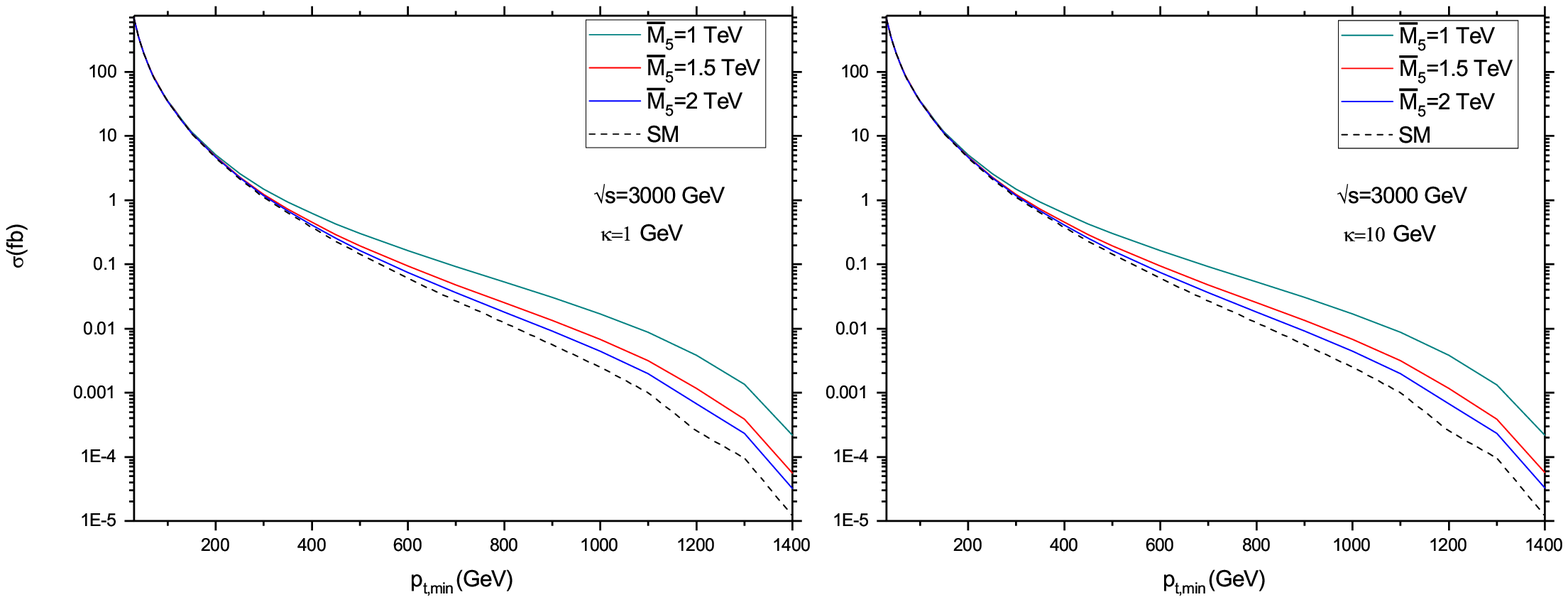}
\caption{The same as in Fig.~\ref{fig:RSSC_750}, but for $\sqrt{s} =
3000$ GeV.} \label{fig:RSSC_1500}
\end{center}
\end{figure}

Finally, we have estimated the total cross-sections for the RS-like
model with the small curvature (RSSC model) using formulas
\eqref{S_RSSC}, \eqref{A_epsilon}. As one can see in
Figs.~\ref{fig:RSSC_750}, \ref{fig:RSSC_1500}, the total cross
sections weakly depend on the curvature parameter $\kappa$ for all
values of the 5-dimensional Planck scale $\bar{M}_5$. It is a
well-known feature of the RSSC model, provided the condition $\kappa
\ll \bar{M}_5$ is satisfied
\cite{Kisselev:2016}-\cite{Kisselev:2006}.
%

\begin{figure}[htb]
\begin{center}
\includegraphics[scale=0.75]{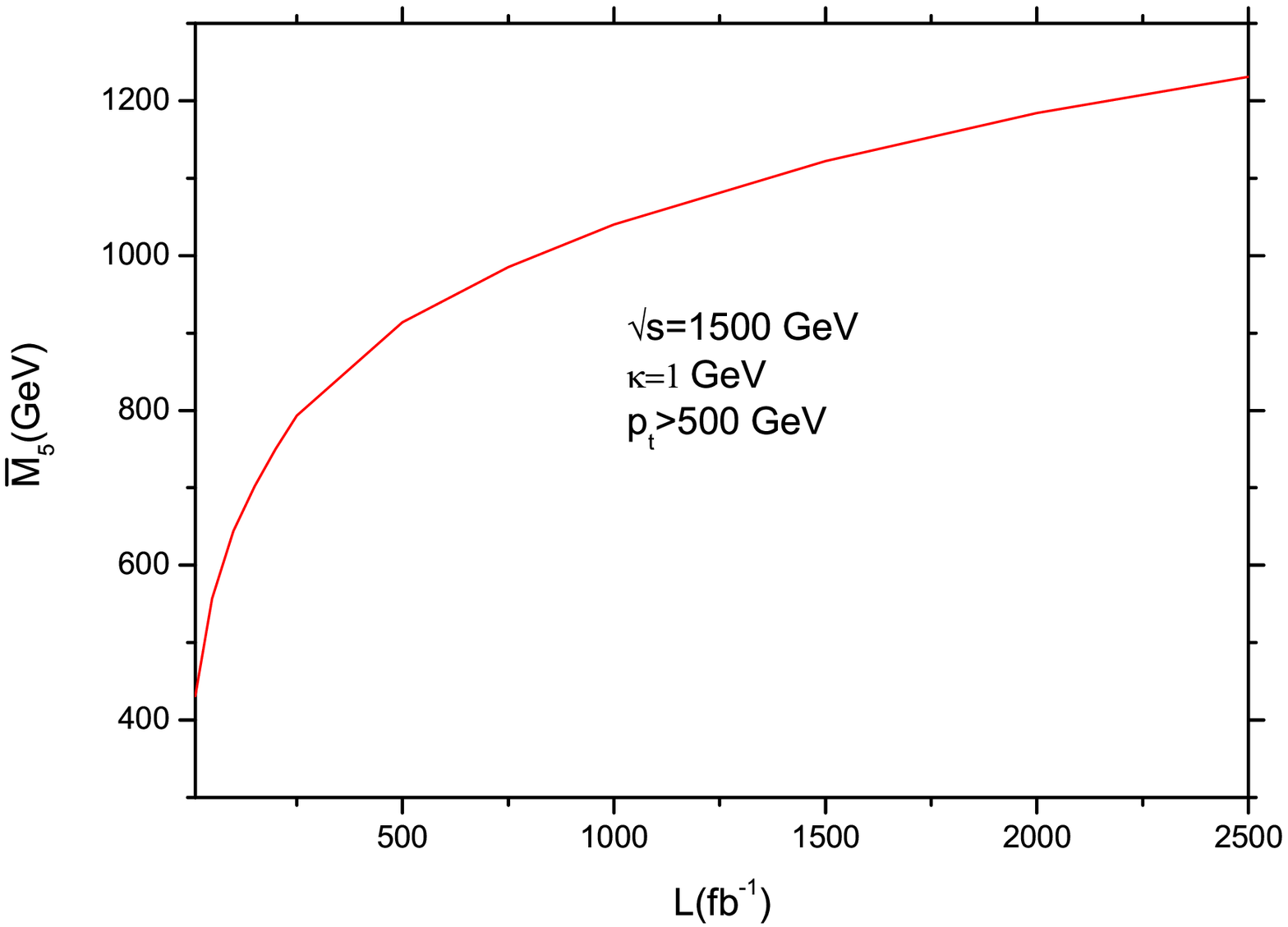}
\caption{The $95\%$ C.L. CLIC search bound in the RSSC model for $\sqrt{s} =
1500$ GeV, $p_t>500$ GeV as a function of the integrated luminosity
$L$.} \label{fig:S_RSSC_750_500}
\end{center}
\end{figure}

\begin{figure}[htb]
\begin{center}
\includegraphics[scale=0.75]{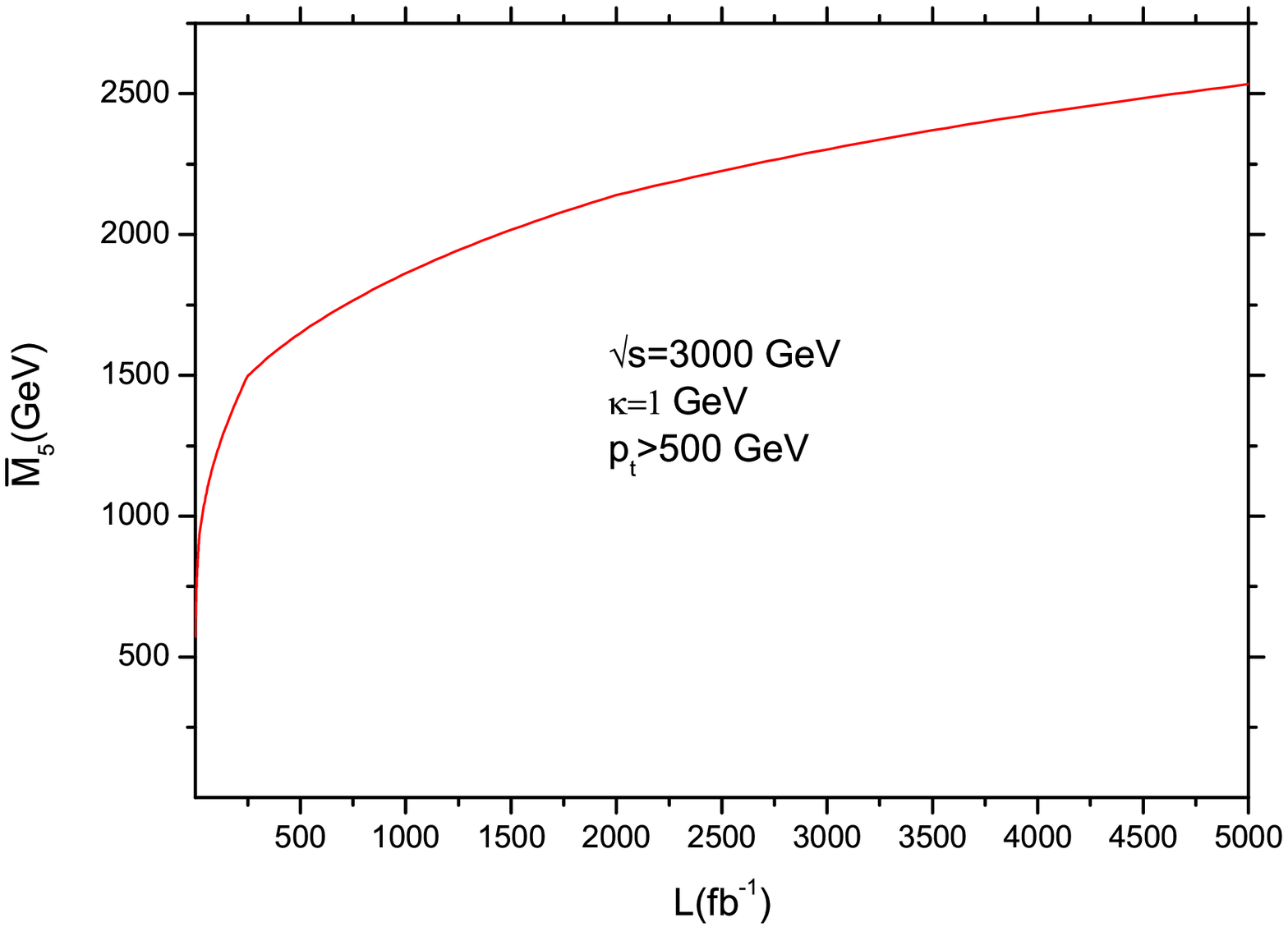}
\caption{The same as in Fig.~\ref{fig:S_RSSC_750_500}, but for
$\sqrt{s} = 3000$.} \label{fig:S_RSSC_1500_500}
\end{center}
\end{figure}

The CLIC search bounds for the scale $\bar{M}_5$ in the RSSC model
are shown in Figs.~\ref{fig:S_RSSC_750_500},
\ref{fig:S_RSSC_1500_500}. The value $\bar{M}_5 = 2534$ TeV can be
achieved for $\sqrt{s} = 3000$ GeV, $L = 5000 \mathrm{\ fb}^{-1}$.
Let us stress that the parameter $\bar{M}_5$ has quite different
magnitudes in the RS and RSSC models. In the RS model, the bounds on
the parameter set ($\beta, m_1$) are searched for. On the contrary,
in the RSSC model one can directly obtain bounds on the
5-dimensional Planck scale $\bar{M}_5$, while dependence on the
curvature parameter $\kappa$ is rather weak. Note that the LHC bound
on $D$-dimensional scale $M_D$ (see, for instance,
\cite{CMS_bounds}) cannot be applied to our lower limits on the
scale $\bar{M}_5$, since the RSSC model cannot be regarded as a
small distortion of the ADD model with one ED \cite{Kisselev:2006}.
%

\section{Conclusions} %

It is expected that LHC gives many answers to some fundamental
problems of particle physics. However, high precision experiments
could not be done at the LHC due to the strong interactions which
spoil the proton structure through the proton-proton interactions.
Therefore, the lepton collider with high luminosity and energy is a
good candidate to complement and develop the LHC physics studies.
Because the CLIC has high energy and luminosity, it is extremely
important to investigate energy-dependent new physics beyond the SM,
such as extra-dimensional models, axion-like particle production,
anomalous gauge couplings, \emph{etc.}

In the present paper we have studied the photon-induced dimuon
production $e^+e^- \rightarrow e^+ \gamma \gamma e^- \rightarrow e^+
\mu^+\mu^- e^-$ at the CLIC in a number of models with EDs. Among
these models are: (i) ADD model with the Han-Lykken-Zhang convention
\cite{Han:1999}; (ii) the ADD model with the Hewett convention
\cite{Hewett:1999}, (iii) the original Randall-Sundrum model
\cite{Randall:1999}, (iv) the Randall-Sundrum-like model with the
small curvature \cite{Giudice:2005}. The total cross sections have
been calculated for energies $\sqrt{s} = 1500$ GeV and $\sqrt{s} =
3000$ GeV. It enabled us to obtain $95\%$ C.L. search limits on the
parameters of the models as functions of the CLIC integrated
luminosity $L$.

In order to suppress the SM background, we imposed the cut on the
transverse momenta of the final muons, $p_t > 500$ GeV. Although the
KK terms dominate the SM term for $p_t > 800$ GeV, our calculations
show that the CLIC sensitivity bounds for the cuts $p_t > 800$ and
$p_t > 500$ GeV are almost the same for small $L$, and differ by
less than 10\% for very high $L$. Note that the number of events
becomes very small for $p_t > 800$. That is why, for the
calculations of the exclusion regions, we used the cut $p_t > 500$
GeV.

The best bounds have been derived for the $e^+e^-$ collision energy
$\sqrt{s} = 3000$ GeV and integrated luminosity $L = 5000 \mathrm{\
fb}^{-1}$. For the ADD model with the HLZ convention, we have
obtained $M_S \geqslant 3629$ GeV (Fig.~\ref{fig:S_HLZ_1500}), while
for the ADD model with the Hewett convention, we have got that $M_H
\geqslant 2204$ GeV (Fig.~\ref{fig:S_H_1500}). The best limits for
the parameters ($\beta, m_1$) of the RS model are presented in
Fig.~\ref{fig:S_RS_1500_500}. The bounds on the fundamental
gravity scale $\bar{M}_5$ in the RSSC model is of considerable
interest, since so far there are no experimental limits on the
parameters of this RS-like model. The best bound on this model parameter
shown in Fig.~\ref{fig:S_RSSC_1500_500}. Note that the LHC discovery limits
on $\bar{M}_5$ for the photon-induced process $pp \rightarrow p
\gamma \gamma p \rightarrow p \mu^+\mu^- p$ have been calculated in
our recent paper \cite{Inan_Kisselev:2018}. The LHC bounds obtained
there are noticeably lower than our CLIC bounds.

The gravity and collider constraints on parameters of the models
with EDs lie in the TeV range. For our calculations, we have used
parameters that are of the order of the current experimental
bounds. Three models we analyzed have different metrics and/or
different physical contents. From this perspective, we have
demonstrated the potentials of CLIC photon-induced reactions for
three different models with EDs.

The searches for effects of new physics at the LHC and CLIC can be
regarded as complementary searches. Let us underline, the great
advantage of the CLIC collider is that it has very clean
backgrounds. Moreover, the CLIC detectors don't need additional
equipment for Weizs\"{a}cker-Williams photon-induced collisions
analyzed in the present paper. That is why we think that studying
such reactions at the CLIC could be one of most important physical
tasks.

It would be interesting to compare our results on the total cross
section and CLIC search limits with the corresponding predictions
for the processes $e^+\gamma \rightarrow e^+ \gamma$ and $e^+e^-
\rightarrow e^+ \gamma \gamma e^- \rightarrow e^+ \mu^+\mu^- e^-$,
where $\gamma$ is the Compton backscattering photon
\cite{Compton_photon}. It will be a subject of our separate
publication.



\section*{Acknowledgments}
This work is partially supported by the
Scientific Research Project Fund  of Sivas Cumhuriyet University
under project number ``F-596''.



\setcounter{equation}{0}
\renewcommand{\theequation}{A.\arabic{equation}}

\section*{Appendix A}
\label{app:A}

\begin{figure}[htb]
\begin{center}
\includegraphics[scale=0.70]{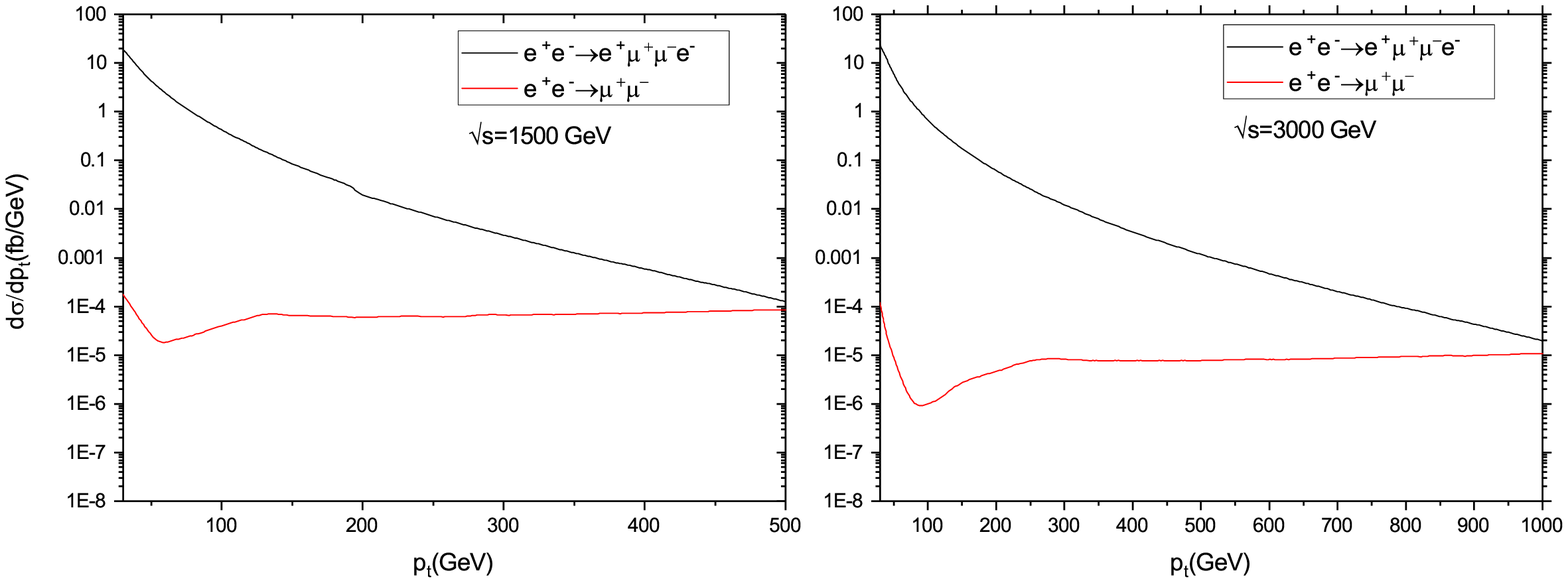}
\caption{The differential cross sections for the processes $e^+e^-
\rightarrow e^+\mu^+\mu^-e^-$ and $e^+e^- \rightarrow \mu^+\mu^-$
for $\sqrt{s} = 1500$ GeV (left panel) and $\sqrt{s} = 3000$ GeV
(right panel).} \label{fig:PTDSMTWOPROCESS}
\end{center}
\end{figure}

One of possible SM backgrounds is the process without electron pair
in the final states. In Fig.~\ref{fig:PTDSMTWOPROCESS} the
differential cross section for the process $e^+e^- \rightarrow
\mu^+\mu^-$ is shown in comparison with that for the SM process
$e^+e^- \rightarrow e^+\mu^+\mu^-e^-$. As one can see, the former is
relatively small up to $p_t = 500 (1000)$ GeV for $\sqrt{s} = 1500
(3000)$ GeV. As a result, with the cut $p_t > 30$ GeV we obtain
$\sigma(e^+e^- \rightarrow \mu^+\mu^-)$ = 6.14 fb and 1.52 fb, for
$\sqrt{s} = 1500$ GeV and $\sqrt{s} = 3000$ GeV, respectively, while
$\sigma(e^+e^- \rightarrow e^+\mu^+\mu^-e^-)$ = 509 fb and 677 fb
for the same energies.

\end{document}